\documentclass[draft,pal]{agutex}

 \usepackage{lineno}
\usepackage{amsmath, amsthm, amssymb}
\usepackage{setspace}
\usepackage{graphicx}
\usepackage[dvips]{epsfig}
\usepackage{paralist}
  \setkeys{Gin}{draft=false}
\authorrunninghead{VIEBAHN ET AL.}

\titlerunninghead{Drake Passage in an eddying global ocean}

\begin{document}

%
%

\title{Effects of Drake Passage on a strongly eddying global ocean}
%
%

%
%



 \authors{Jan P. Viebahn,\altaffilmark{1}
 Anna S. von der Heydt,\altaffilmark{1} Dewi Le Bars,\altaffilmark{1}
 and Henk A. Dijkstra\altaffilmark{1}}

\altaffiltext{1}{Institute for Marine and Atmospheric Research Utrecht,
Utrecht University, Utrecht, Netherlands.}





%
%


\begin{abstract}
The climate impact of ocean gateway openings during the Eocene-Oligocene 
transition is still under debate.
Previous model studies employed 
grid resolutions at which
the impact of mesoscale eddies has to be parameterized.
We present results of a state-of-the-art eddy-resolving global ocean model
with a closed Drake Passage, and compare with results of the same model at non-eddying resolution.
An analysis of the pathways of heat by decomposing the meridional
heat transport
into eddy, horizontal, and overturning circulation components
indicates that the model behavior on the large scale is qualitatively similar at both resolutions.
Closing Drake Passage induces
$(i)$ sea surface warming around Antarctica 
due to changes in the horizontal circulation of the Southern Ocean,
$(ii)$ the collapse of the overturning circulation related to North Atlantic Deep Water formation
leading to surface cooling in the North Atlantic,
$(iii)$ significant equatorward eddy heat transport near Antarctica.
However, quantitative details significantly depend on the chosen resolution.
The warming around Antarctica is substantially larger 
for the non-eddying configuration ($\sim$$5.5^\circ$C)
than for the eddying configuration ($\sim$$2.5^\circ$C).
This is a consequence of the subpolar mean flow which
partitions differently into gyres and circumpolar current at different resolutions.
We conclude that 
for
a deciphering of the different mechanisms active in 
Eocene-Oligocene climate change
detailed 
analyses of the pathways of heat 
in the different climate subsystems
are crucial 
in order to clearly 
identify the 
physical processes actually at work. 

\end{abstract}

%
%

%

\begin{article}

%
%

\section{Introduction}
During the past 65 Million years (Ma), climate has undergone a major change from a
warm and ice-free ``greenhouse" to colder ``icehouse" conditions with extensive continental ice sheets and polar ice caps \citep{Katz:08}.
The Eocene-Oligocene boundary ($\sim$34 Ma) reflects a major transition
and the first clear step into icehouse conditions during the Cenozoic.
It is characterized by a rapid expansion of large permanent continental ice sheets on Antarctica 
which is superimposed on a long-term gradual cooling trend in Cenozoic global climate change \citep{Zachos:01,Zachos:08}.

Proposed mechanisms of the onset of Oligocene glaciation 
include, on the one hand,
increased thermal isolation of Antarctica due to the reorganization of the global ocean circulation induced by critical tectonic opening/widening of ocean gateways surrounding Antarctica \citep{Kennett:77,Toggweiler:00,Exon:02,Livermore:07,Yang:13}.
We refer to this perspective as the \emph{ocean gateway mechanism}.
On the other hand, declining atmospheric $CO_2$ concentration from peak levels in the early Eocene has been
suggested as a dominant process inducing Antarctic ice sheet growth, the so-called \emph{$CO_2$ drawdown mechanism} \citep{DeConto:03,Pagani:11}.

Recently, there has been a tendency to discuss these different mechanisms
as hypotheses competing against each other \citep{Goldner:14}.
However, a dynamical scenario with several mechanisms at work
which can trigger 
or feedback on each other might ultimately more adequately capture the 
nonlinearity of climate dynamics \citep{Livermore:05,Scher:06,Katz:08,Miller:09,Sijp:09,Lefebvre:12,Dijkstra:13}.
Unfortunately, revealing 
the precise relative roles of different mechanisms 
active in Cenozoic climate change
via comprehensive climate model sensitivity studies is still extremely limited \citep{Wunsch:10}. 
Crucial initial and boundary conditions are highly uncertain due to the limited amount of proxy data.
In addition, the accurate resolution of the broad range of spatial and temporal scales involved (ranging from fine topographic structures both in the ocean and on Antarctica to the long time scales of continental ice sheet dynamics)
exceeds current computational resources.

Correspondingly, the actual physical processes at work within a single mechanism are still under debate \citep{DeConto:07}.
Regarding the ocean gateway mechanism it is far from clear
how the changes in poleward ocean heat transport are dynamically accomplished \citep{Huber:06}.
On the one hand,
polar cooling was related to changes in the horizontal gyre circulation
with the focus being on either the subtropical western boundary circulation \citep{Kennett:77,Exon:02}
or the subpolar poleward flow along eastern boundaries \citep{Huber:04,Huber:06,Sijp:11}.
On the other hand, decreased poleward ocean heat transport
was generally attributed to changes in the meridional overturning circulation (MOC) \citep{Toggweiler:00,Sijp:04,Sijp:09,Cristini:12,Yang:13}.
The MOC, however,
can be decomposed into shallow wind-driven gyre overturning cells,
deep adiabatic overturning cells (related to North Atlantic Deep Water (NADW) formation),
and deep diffusively driven overturning cells (related to bottom waters) \citep{Kuhlbrodt:07,Ferrari:11,Wolfe:14}.
In other words, the actual pathways of heat and the related physical processes have not been
precisely analyzed in previous paleoceanographic model studies
(the most extensive studies in this respect are \cite{Huber:06} and \cite{Sijp:11}).

Next to the large-scale circulation regimes this also concerns the impact of mesoscale eddies
which are known to be of leading-order importance for both the ocean mean state and its response to changing forcing, especially in the Southern Ocean (SO) \citep{Hallberg:06,Munday:15}.
For example, interfacial eddy form stresses are crucial in transporting the momentum inserted by the surface wind stress
downward to the bottom where they are balanced by bottom form stresses
(i.e. the interaction between pressure and topography) \citep{Munk:51,Olbers:04,Ward:11}.
Moreover, interfacial eddy form stresses are equivalent to southward eddy heat fluxes \citep{Bryden:79},
which represent a significant contribution to the total heat flux in specific regions \citep{Jayne:02,Meijers:07}.
For more details on the crucial role of eddies in both the zonal circulation and the meridional overturning circulation
we refer to \cite{Rintoul:01} and \cite{OWE:12}.

Previous studies on the climatic impact of ocean gateways using general circulation models vary in complexity, including ocean-only, ocean with simple atmosphere, and more recently, coupled ocean-atmosphere models (for an overview see Table 1 in \cite{Yang:13}). However, all these models are based on rather coarse ocean model grid resolutions, with the highest horizontal resolution being larger than $2^\circ$.
As computer power has increased over the past decade, many climate models used for projections of future climate change operate nowadays with a horizontal resolution of about $1^\circ$ in the ocean component \citep{IPCC:13}. 
But even such a resolution is still too coarse to admit mesoscale eddies.
Consequently, some version of the Redi neutral diffusion \citep{Redi:82} and Gent and McWilliams (GM) eddy advection parameterization \citep{Gent:90} is typically employed in order to represent the effects of mesoscale eddies.

The computational cost of eddy-resolving resolutions is still so high that coupled climate model simulations with resolved eddies have only recently been attempted with relatively short integration times \citep{McClean:11,Kirtman:12}.
The required huge computational resources eliminate the possibility of multiple control simulations and, hence, the corresponding tuning of model parameters. This also renders the comparison of high-resolution eddy-resolving climate models with low-resolution non-eddy-resolving climate models difficult, because the control run climatologies of models at different resolutions are usually quite different.
Consequently, investigations of how well coupled climate models using mesoscale eddy parameterizations are able to represent the current climate state as well as
the response to changes in climate forcing/boundary conditions are only beginning to emerge \citep{Bryan:14,Griffies:15}.

For ocean-only models eddy-resolving horizontal resolutions of about $0.1^\circ$
are nowadays possible even for global-scale simulations.
There is no doubt that such a resolution is able to give a much more realistic representation of mesoscale eddies and strong and narrow currents than when coarser resolutions are used \citep{Farneti:10,Farneti:11,Gent:11}.
In this study, for the first time, the effects of Drake Passage (DP) in both high-resolution eddy-resolving global ocean model experiments (i.e. nominal $0.1^\circ$) and low-resolution non-eddy-resolving experiments (i.e. nominal $1^\circ$)
are compared.
In line with many previous studies \citep{Mikolajewicz:93,Sijp:04,Sijp:09,Yang:13}, we isolate the influence of DP by closing DP with a land bridge and keeping all other boundary conditions as at present.
Of course, ocean-only models miss the interactions with other components of the climate system.
On the other hand, the superposition of multiple feedbacks with additional changes in boundary conditions,
such as poorly-constrained changes of atmospheric carbon dioxide, can obscure the specific effects of the gateways.
In other words, the effects of gateways themselves can be most clearly assessed in relatively simple experiments
for which only the gateways are changed while all other boundary conditions are held equal \citep{Yang:13}.
This way our high-resolution ocean-only simulations may serve as a reference for upcoming high-resolution coupled climate model simulations as well as high-resolution simulations with paleoclimate topographies.
In particular, we employ 
the ocean component of the Community Earth System Model (CESM) in this study, a climate model that is generally well assessed among the climate models participating in CMIP5 \citep{IPCC:13},
and that has already been analyzed in low-resolution paleoconfigurations \citep{Huber:04,Huber:06}.
This will allow for comparisons with both previous high-resolution present-day simulations (e.g. \cite{Bryan:14}) and future high-resolution paleoclimate simulations employing CESM.

The layout of the paper is as follows:
In section \ref{model} we describe the details of the model configuration and the simulations.
In sections \ref{flowfield}-\ref{MHT_sec} we compare eddying model simulations and non-eddying model simulations
with both open and closed DP in terms of 
changes in flow field, SST, and meridional heat transport respectively. 
Section \ref{discussion} provides a summary and discussion,
and we close with conclusions and perspectives in section \ref{conclusions}.

\section{Model and simulations}\label{model}
The global ocean model simulations analyzed in this study are performed with the Parallel Ocean Program (POP, \cite{Dukowicz:94}), developed at Los Alamos National Laboratory.
We consider the same two POP configurations as in \cite{Weijer:12} and \cite{DenToom:14}.
The strongly-eddying configuration, indicated by $R0.1$, has a nominal horizontal resolution of $0.1^\circ$,
which allows the explicit representation of energetic mesoscale features including eddies and boundary currents \citep{Maltrud:10}.
The lower-resolution (``non-eddying'') configuration of POP, indicated by $x1$, has the nominal $1.0^\circ$ horizontal resolution. 
The two versions of the model are configured to be consistent with each other, where possible.
There are, however, some notable differences, which are discussed in full in the supplementary
material of \cite{Weijer:12}. 
For the present study we note that the $R0.1$ configuration has a tripolar 
grid layout, with poles in Canada and Russia, whereas the $x1$ configuration is based on a dipolar grid, with the northern pole displaced onto Greenland. In the $R0.1$ configuration, the model has $42$ non-equidistant z-levels, increasing in thickness from $10\ m$ just below the upper boundary to $250\ m$ just above the lower boundary at $6000\ m$ depth.
In addition, bottom topography is discretized using partial bottom cells, creating a more accurate and smoother representation of topographic slopes.
In contrast, in the $x1$ configuration the bottom is placed at $5500\ m$ depth, there are $40$ levels (with the same spacing as in $R0.1$), and the partial bottom cell approach is not used.
Finally, in the $x1$ configuration tracer diffusion is accomplished by the GM eddy transport scheme \citep{Gent:90}
using a constant eddy diffusivity of $600\ m^2/s$ (and tapering towards the surface).
In summary, the two POP configurations employ different representations of both
mesoscale eddies and topographic details which consequently may lead
to different interactions between mean flow, mesoscale eddies and topography \citep{Adcock:00,Dewar:02,Barnier:06,Sommer:09}.

The atmospheric forcing of the model is based on the repeat annual cycle (normal-year) Coordinated Ocean Reference Experiment (CORE\footnote{http://www.clivar.org/organization/wgomd/core}) forcing dataset \citep{Large:04}, with 6-hourly forcing averaged to monthly. Wind stress is computed offline using the Hurrell Sea Surface Temperature (SST) climatology \citep{Hurrell:08} and standard bulk formulae; evaporation and sensible heat flux were calculated online also using bulk formulae and the model predicted SST. Precipitation was also taken from the CORE forcing dataset. Sea-ice cover was prescribed based on the $-1.8^\circ C$ isoline of the SST climatology, with both temperature and salinity restored on a timescale of 30 days under diagnosed climatological sea-ice.

As initial condition for the $R0.1$ simulations we use the final state of a 75 years spin-up simulation described in \cite{Maltrud:10} using restoring conditions for salinity. The freshwater flux was diagnosed during the last five years of this spin-up simulation and the simulations we present in this paper use this diagnosed freshwater flux to avoid restoring salinity (see also \cite{Weijer:12} and \cite{DenToom:14}). Using mixed boundary conditions makes the ocean circulation more free to adapt, in particular, to changes in continental geometry \citep{Sijp:04}.
For the $x1$ configuration the same procedure as for $R0.1$ is followed.
Additionally, for $x1$ a more equilibrated state is obtained by integrating the model for an additional 500 years under mixed boundary conditions. 
The $x1$ simulations then branch off from this state (i.e. end of year 575).

For both 
configurations, we utilize results of the following two simulations: the unperturbed reference
simulation with a present-day bathymetry ($\mathrm{DP_{op}}$), and a simulation in which DP is closed by a land bridge between the Antarctic Peninsula and South America ($\mathrm{DP_{cl}}$). 
Apart from a closed DP, the $\mathrm{DP_{cl}}$ simulations are conducted with the same present-day bathymetry as for the corresponding $\mathrm{DP_{op}}$ simulations.
The $\mathrm{DP^{R0.1}_{cl}}$ simulation has a duration of 200 years such that the strong adjustment process 
due to closing DP is largely past.
In contrast, the $\mathrm{DP^{x1}_{cl}}$ experiment was carried out for a total of 1000 years such that it is closer to a completely equilibrated state.
The $\mathrm{DP_{op}}$ simulations are also 
analyzed in \cite{Weijer:12}, \cite{DenToom:14}, and
\cite{LeBars:15}, with \cite{Weijer:12} presenting a comparison with observations in their auxiliary material.
Here we use the $\mathrm{DP_{op}}$ simulations in order to 
demonstrate the effects of closing DP.
For that reason we compare each $\mathrm{DP_{cl}}$ simulation with the annual average of the initial year of the corresponding reference simulation (i.e. year $76$ for $\mathrm{DP^{R0.1}_{op}}$ and year $576$ for $\mathrm{DP^{x1}_{op}}$). For the eddy heat transport we additionally consider multidecadal averages of the initial 20 years of the $\mathrm{DP_{op}}$ simulations.
We note that quantitative differences which would appear if later years or longer averages are used (i.e. due to low-frequency internal variability or a small trend related to model drift) are small compared to the changes induced by closing DP.



%
%

\newpage
\section{Flow field evolution}\label{flowfield}
For both model configurations $R0.1$ and $x1$ closing DP leads to two major reorganizations of the ocean circulation, namely, the integration of the former Antarctic Circumpolar Current (ACC) into the subpolar gyre system (section \ref{barotropic_sec}) and the shutdown of NADW formation (section \ref{overturning_sec}).

\subsection{Barotropic circulation}\label{barotropic_sec}
\begin{figure}
 \noindent\includegraphics[height=20pc]{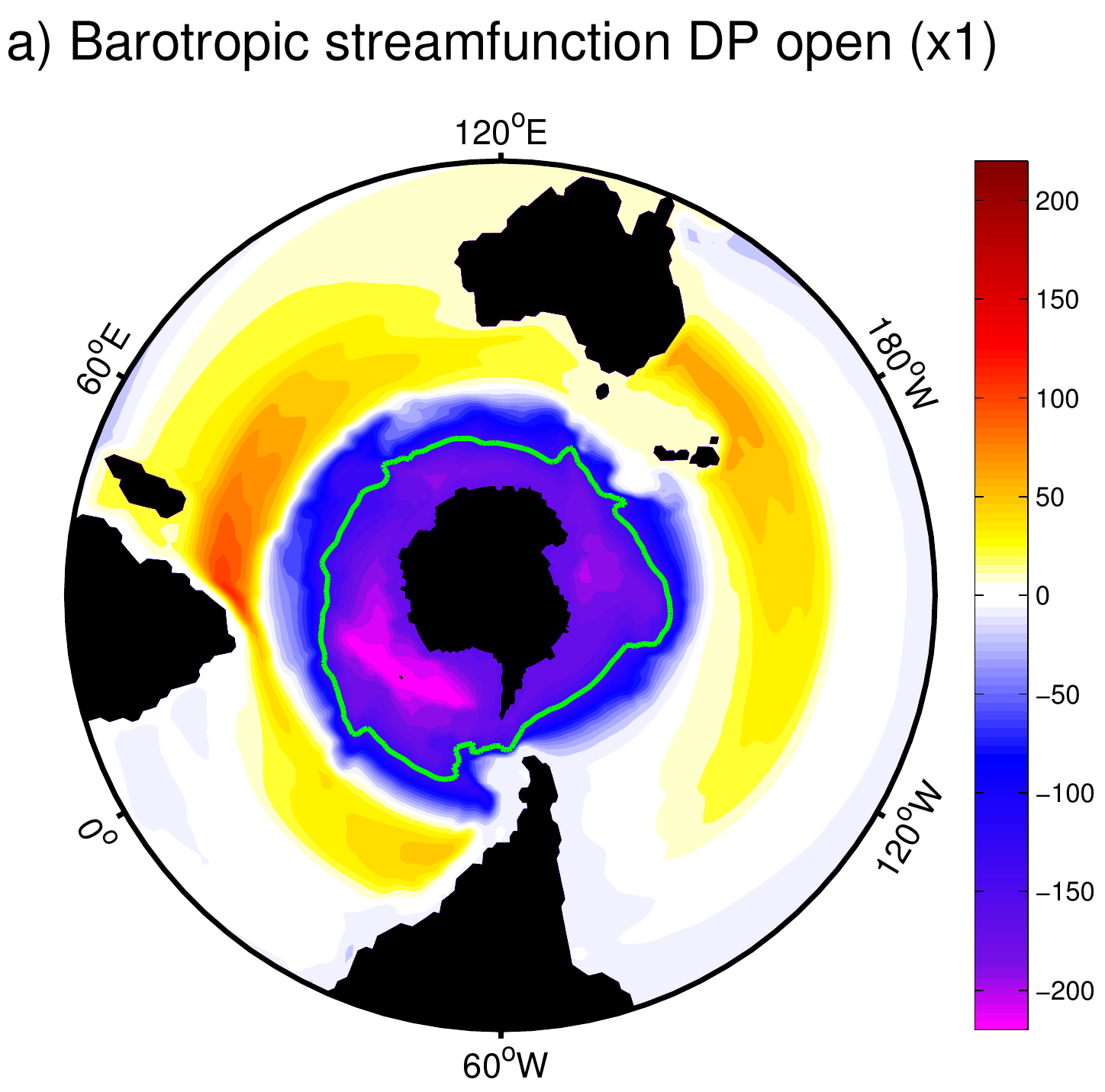}
 \noindent\includegraphics[height=20.pc]{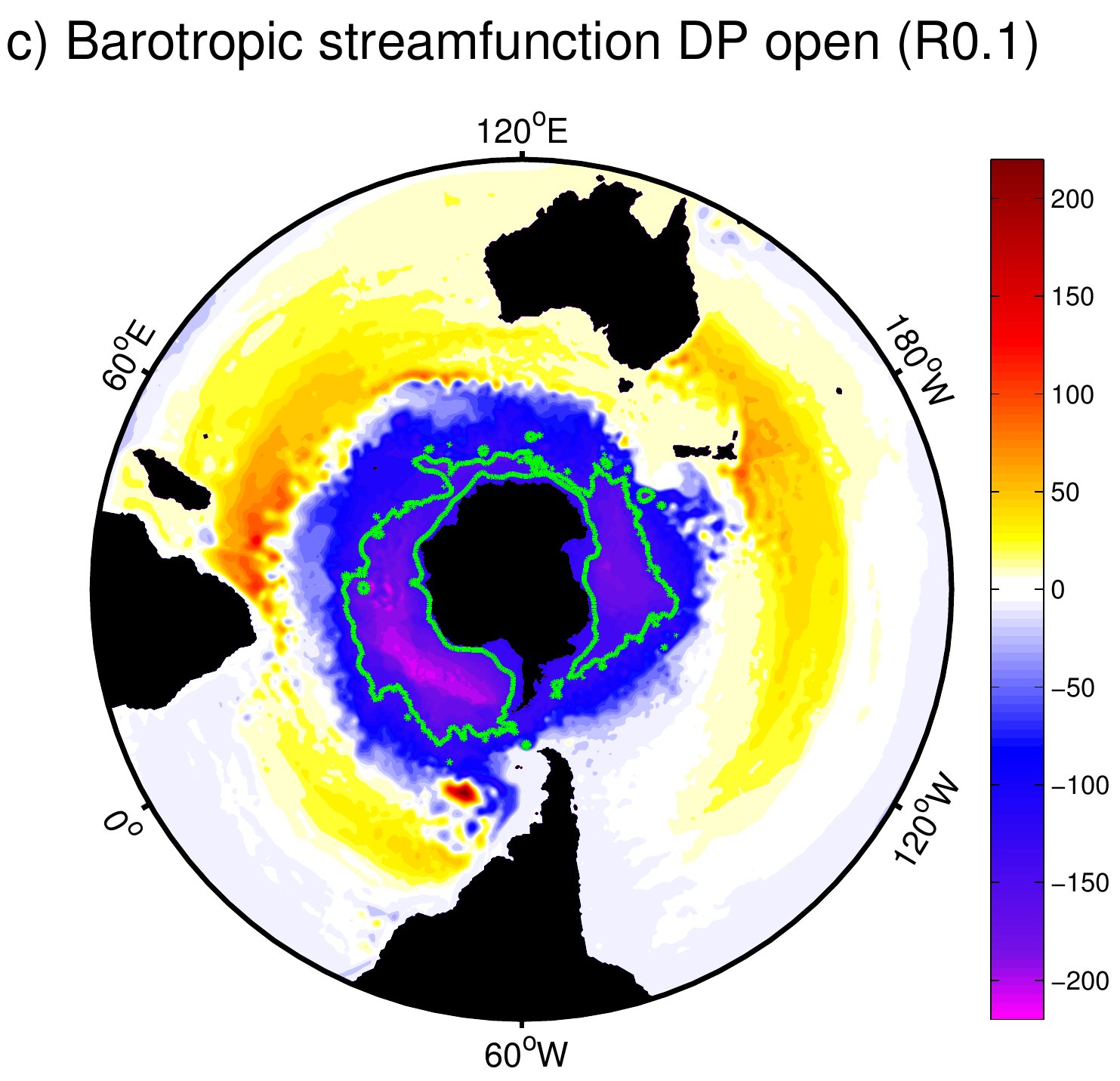}
 \noindent\includegraphics[width=20.5pc]{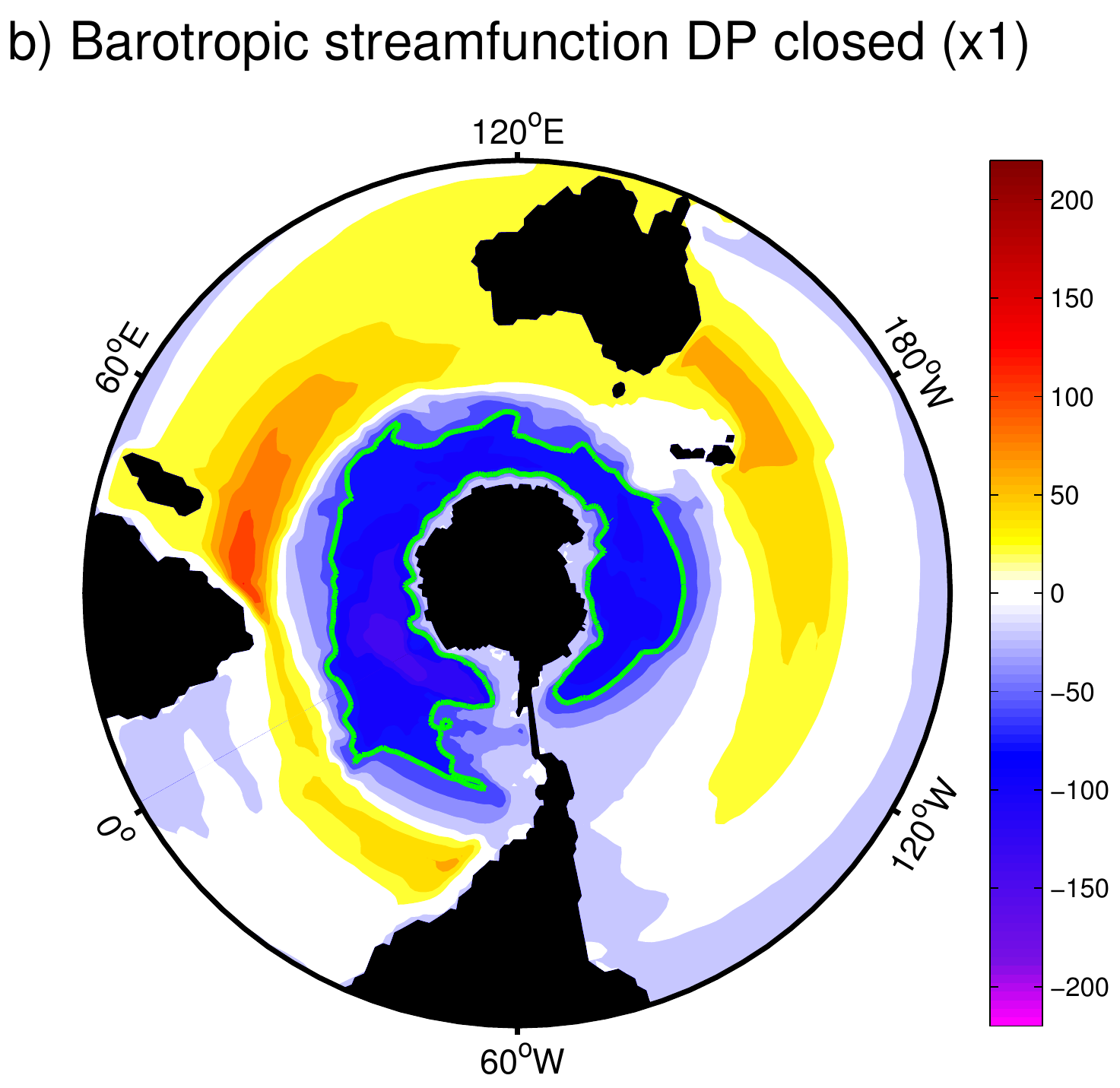}
 \noindent\includegraphics[height=20pc]{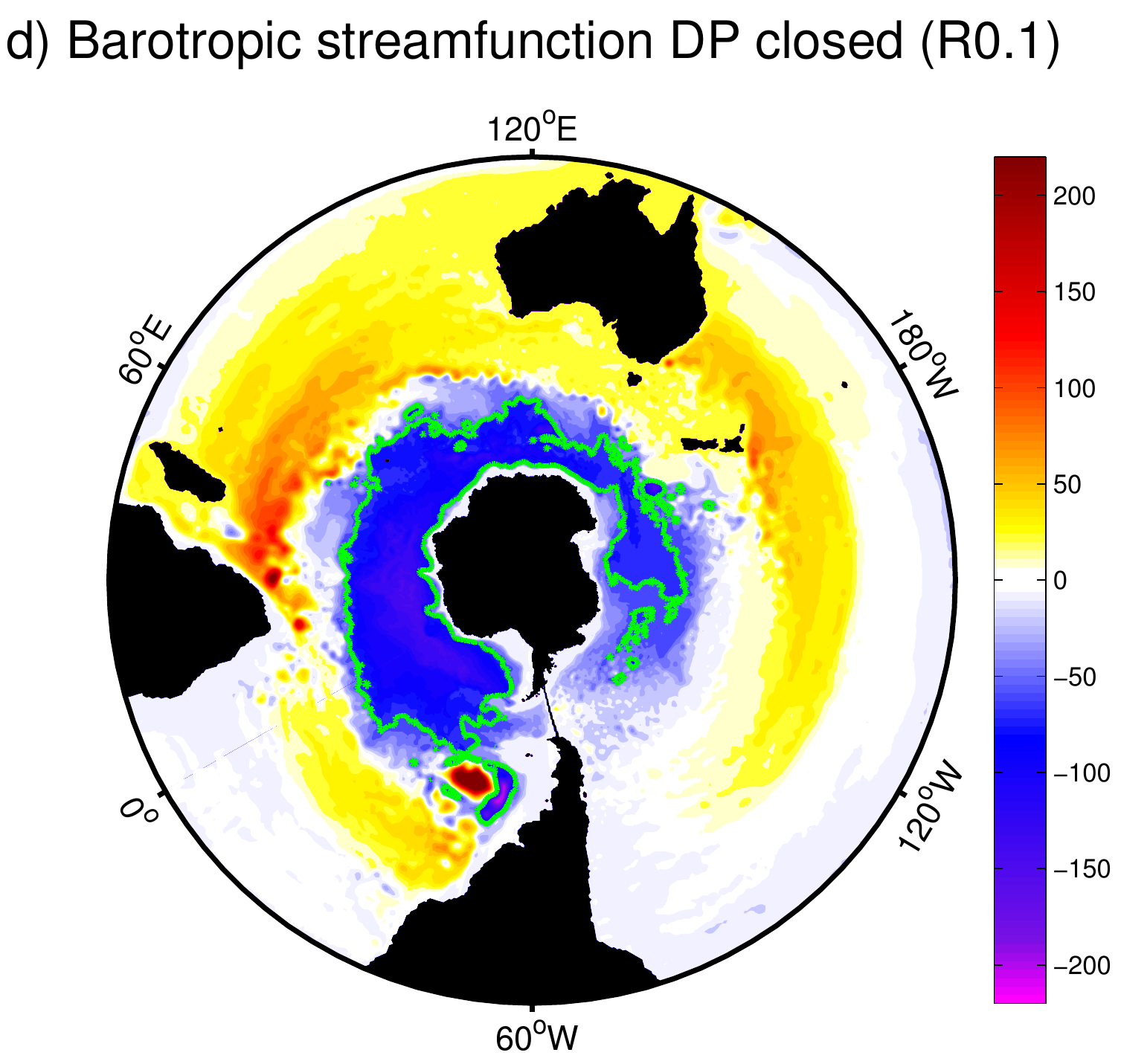}
\caption{Barotropic streamfunction [Sv] for low resolution (a) DP open and (b) DP closed, and high resolution (c) DP open and (d) DP closed. The green lines highlight contours of (a,c) -130 Sv and (b,d) -60 Sv.
Shown is the annual average of year (a) 576, (b) 775, (c) 76, (d) 275.}
\label{barotropic}
\end{figure}

Figure \ref{barotropic} shows the annual-mean barotropic streamfunction in the SO for $\mathrm{DP^{x1}_{op}}$ (Fig. \ref{barotropic}a) and $\mathrm{DP^{R0.1}_{op}}$ (Fig. \ref{barotropic}c).
The overall large-scale patterns are qualitatively similar and consist of three major regimes:
a subtropical gyre within each ocean basin extending to about $40^\circ S$,
the ACC passing through DP and performing large meridional excursions steered by topography \citep{Rintoul:01,Olbers:04},
and the Weddell and Ross subpolar gyres in the Atlantic basin and Pacific basin respectively.
However, the two mean states differ in DP transport
($159\ Sv$ for $\mathrm{DP^{x1}_{op}}$ and $119\ Sv$ for $\mathrm{DP^{R0.1}_{op}}$)
as well as maximum westward Weddell gyre transport
($-60\ Sv$ for $\mathrm{DP^{x1}_{op}}$ and $-90\ Sv$ for $\mathrm{DP^{R0.1}_{op}}$).
That is, the overall streamfunction minimum (located around $61^\circ S$) of the two mean states is similar (about $10\ Sv$ difference),
whereas the partitioning of the subpolar flow into circumpolar current and subpolar gyres is 
quantitatively substantially different.
This is indicated in Fig. \ref{barotropic}a,c by the green lines showing the contours related to $-130\ Sv$ 
for both $\mathrm{DP_{op}}$ cases.

Moreover, on top of the large-scale pattern finer eddying structures are present for $\mathrm{DP^{R0.1}_{op}}$ (Fig. \ref{barotropic}c).
In the western boundary currents and along strong frontal regions of the ACC
mixed barotropic-baroclinic instabilities induce smaller-scale temporal and spatial variability
including mesoscale eddies. 
The most noticeable local features are propagating eddies in the Agulhas Current retroflection region
\citep{Biastoch:08,LeBars:14}
and the bipolar mode close to Zapiola Rise in the Argentine Basin \citep{Weijer:07}.
We also note that the quantitative details of both the pathway and transport of the circumpolar current 
depend on,
on the one hand, interfacial eddy form stresses which transport momentum vertically \citep{Rintoul:01,Olbers:01,Ward:11,Nadeau:15},
and on the other hand, the topographic details which are part (via topography-eddy-mean flow interactions) of the ultimate balance between input of momentum by surface winds and bottom form stress \citep{Olbers:04,Hogg:14,Munday:15}.

Figure \ref{barotropic} also shows the annual-mean barotropic streamfunction in the SO for $\mathrm{DP^{x1}_{cl}}$ (Fig. \ref{barotropic}b) and $\mathrm{DP^{R0.1}_{cl}}$ (Fig. \ref{barotropic}d).
Again the overall large-scale circulation patterns are similar and now consist of two major regimes.
The former ACC streamlines are mostly 
integrated into the subpolar gyre system
such that the westward subpolar gyre transports drastically increase.
For example, for both configurations ($x1$ and $R0.1$) the westward Weddell gyre transport initially increases to about $200\ Sv$,
which subsequently reduces to about $150\ Sv$ during the adjustment process (not shown)
such that the overall streamfunction minimum in Fig. \ref{barotropic}b,d is smaller compared to the corresponding $\mathrm{DP_{op}}$ cases (Fig. \ref{barotropic}a,c). The green lines in Fig. \ref{barotropic}b,d show the contours related $-60\ Sv$ 
and indicate the overall similarity of the subpolar flow fields of $\mathrm{DP^{x1}_{cl}}$ and $\mathrm{DP^{R0.1}_{cl}}$.
On the other hand, the large-scale subtropical gyre system remains relatively unaffected by closing DP (for both $x1$ and $R0.1$)
with only a small part of the former ACC transport entering (the Indonesian throughflow and Mozambique Current transports increase by about $5\ Sv$).

Consequently, the 
differences between the configurations $x1$ and $R0.1$ in the mean barotropic circulation 
appear to be less drastic for $\mathrm{DP_{cl}}$ (Fig. \ref{barotropic}b,d) than for $\mathrm{DP_{op}}$ (Fig. \ref{barotropic}a,c).
Of course, the explicit representation of mesoscale instabilities in $\mathrm{DP^{R0.1}_{cl}}$ (Fig. \ref{barotropic}d)
induces smaller-scale spatial and temporal variability which is not present in $\mathrm{DP^{x1}_{cl}}$ (Fig. \ref{barotropic}b). 
But closing DP suspends the issue of the partitioning of the subpolar flow into circumpolar flow and subpolar gyres.
This makes the flow dynamics less intricate and potentially less prone to differences in resolution.



\subsection{Meridional Overturning circulation}\label{overturning_sec}
\begin{figure}
 \noindent\includegraphics[width=20pc]{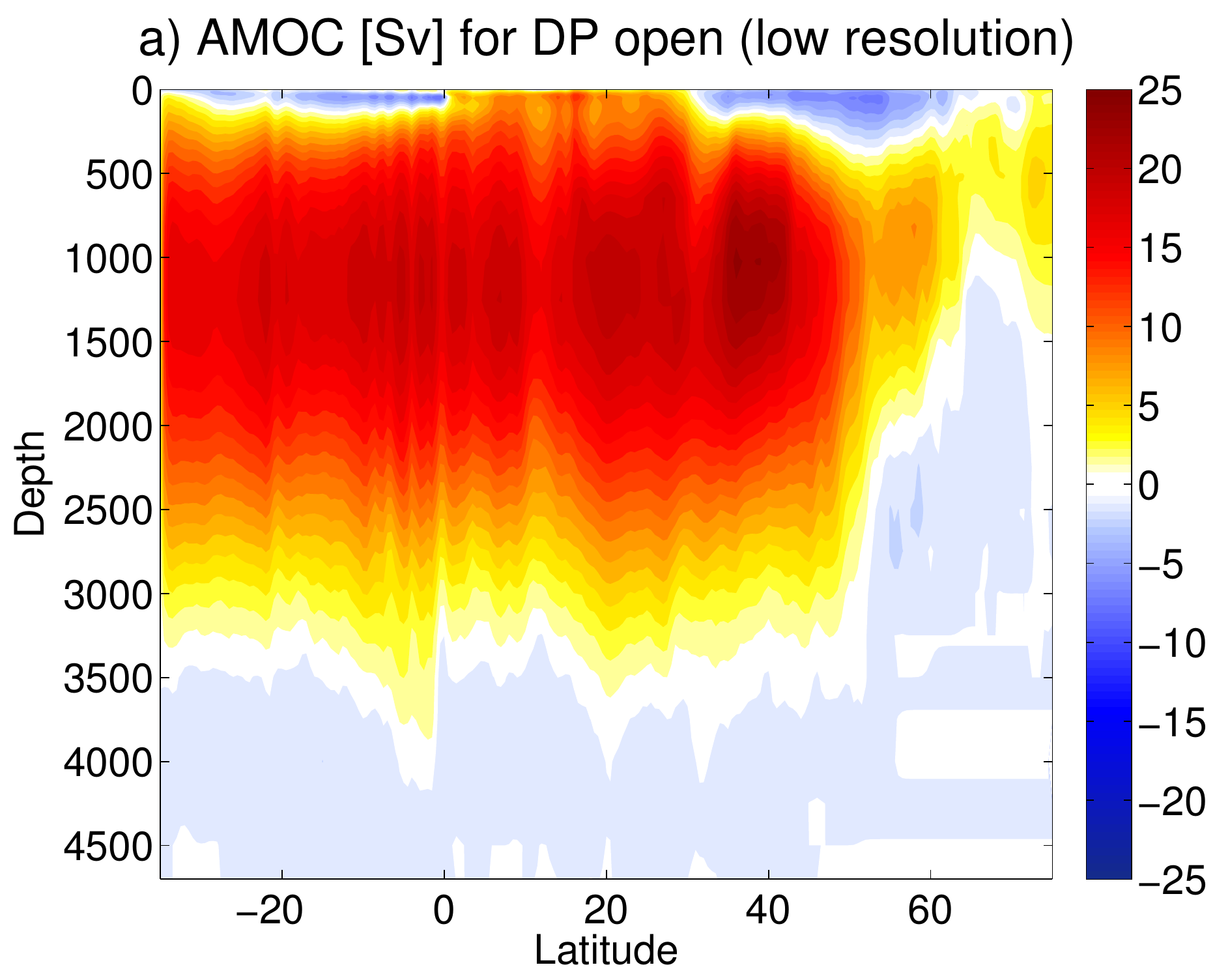}
 \noindent\includegraphics[width=20pc]{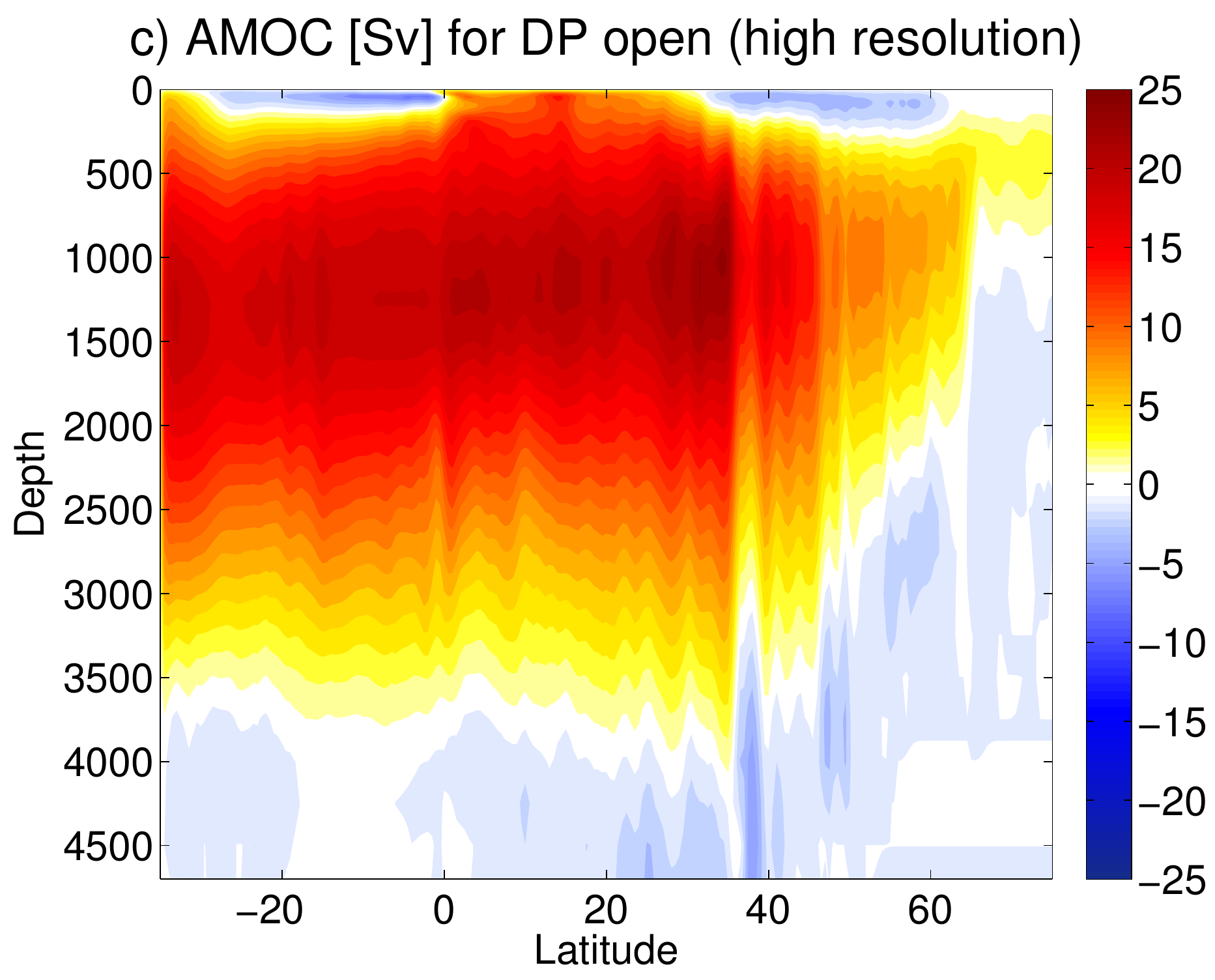}
 \noindent\includegraphics[width=20pc]{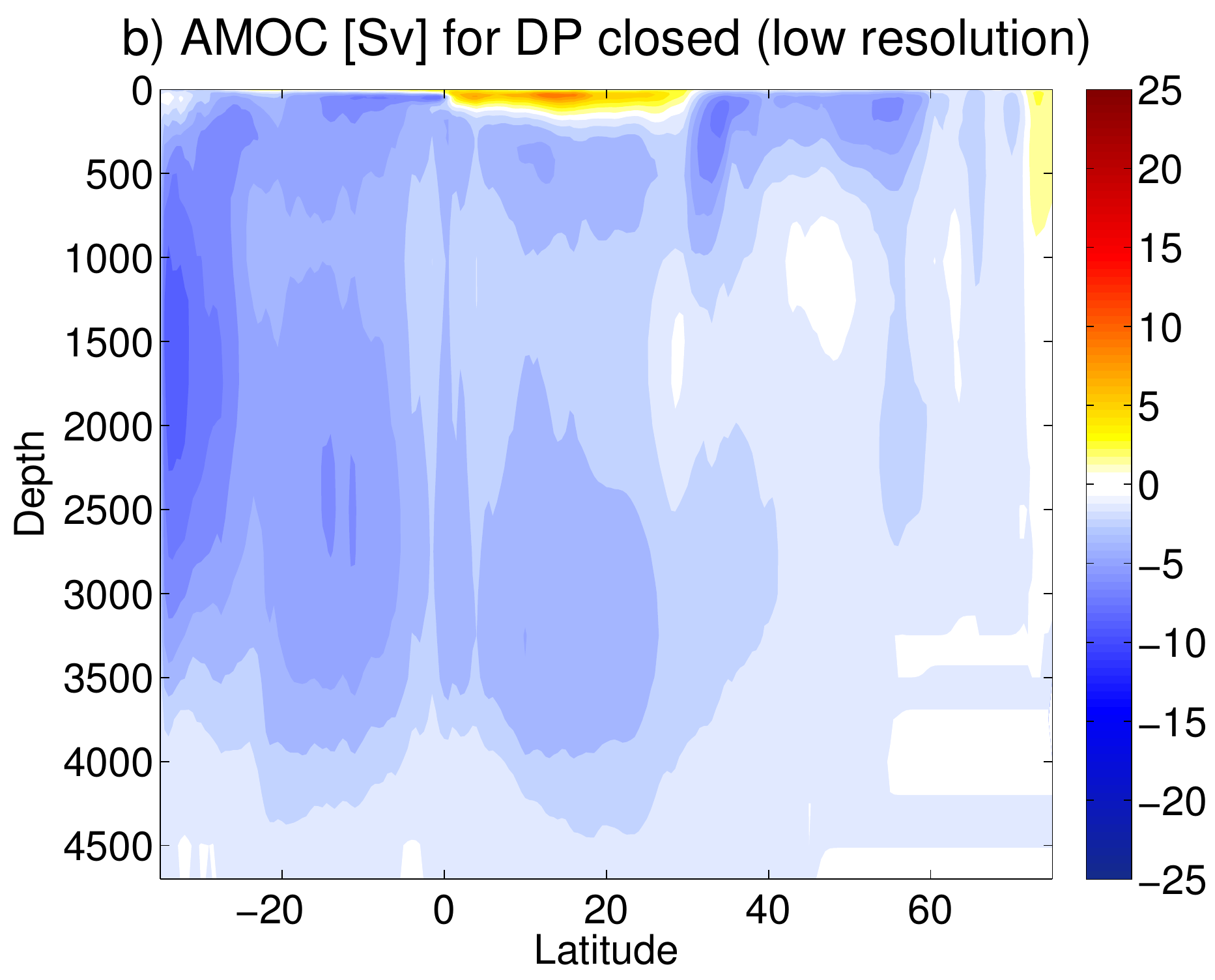}
 \noindent\includegraphics[width=20pc]{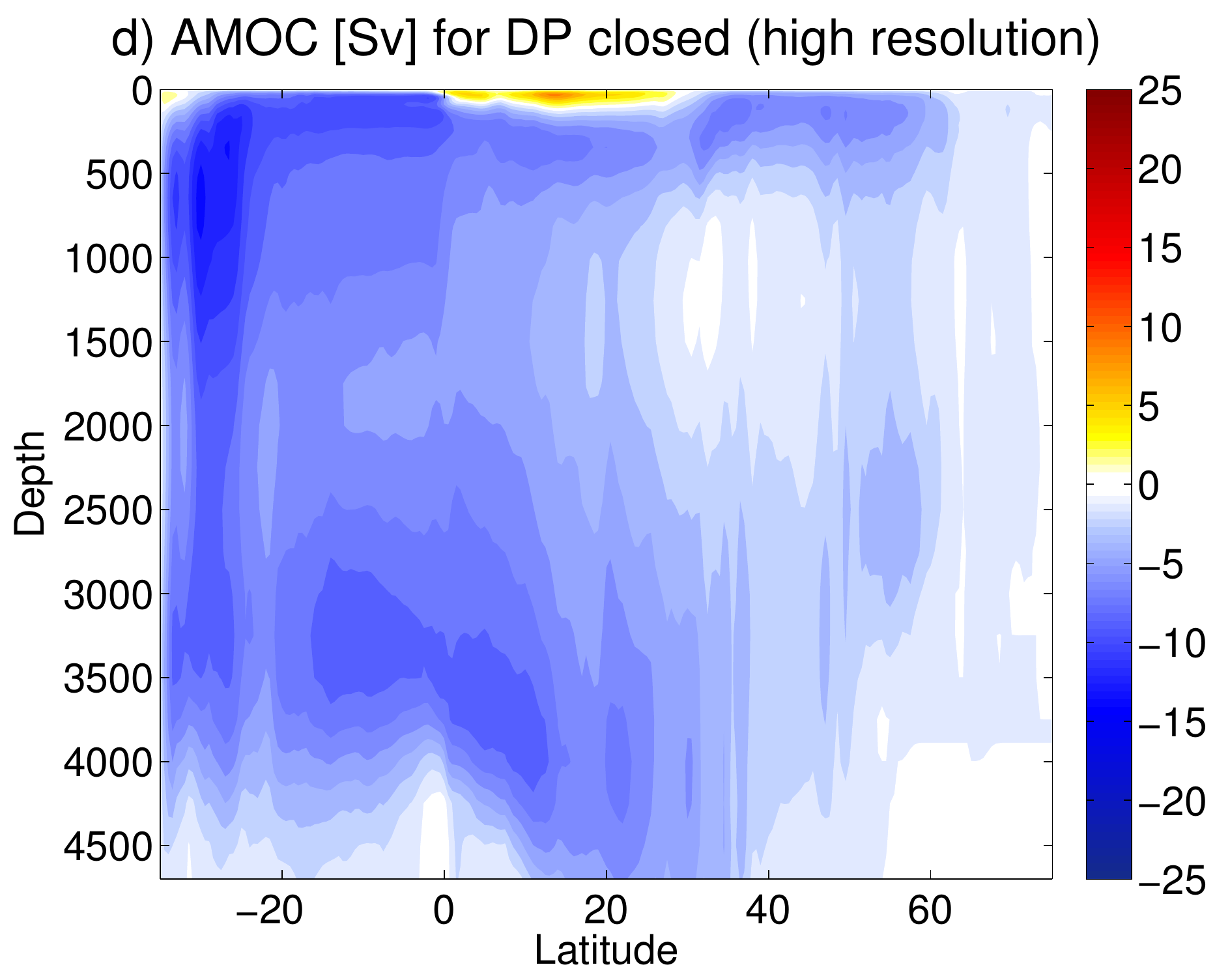}
\caption{Atlantic meridional overturning streamfunction [Sv] for low resolution (a) DP open and (b) DP closed, and high resolution (c) DP open and (d) DP closed. Shown is the annual average of year (a) 576, (b) 775, (c) 76, (d) 275. }
\label{AMOC}
\end{figure}

Figure \ref{AMOC} shows the annual-mean Atlantic MOC (AMOC) for all model simulations.
The AMOC patterns of $\mathrm{DP^{x1}_{op}}$ (Fig. \ref{AMOC}a) and $\mathrm{DP^{R0.1}_{op}}$ (Fig. \ref{AMOC}c) are very similar to each other (see also \cite{Weijer:12}).
They exhibit the typical present-day Atlantic upper overturning cell with northward flow near the surface (interlooped with shallow-gyre overturning cells), NADW formation in the subpolar regions (seas surrounding Greenland),
and southward flow at mid-depth (along the western boundary of the Atlantic Ocean).
For $\mathrm{DP^{x1}_{op}}$ ($\mathrm{DP^{R0.1}_{op}}$) the maximum of the AMOC is $24.0$~$Sv$ ($24.1$~$Sv$) and centered at $36^\circ N$ ($34.5^\circ N$) and about 1000 $m$ depth.
The deep overturning cell related to Antarctic Bottom Water is very weak.

In contrast, for both $\mathrm{DP^{x1}_{cl}}$ (Fig. \ref{AMOC}b) and $\mathrm{DP^{R0.1}_{cl}}$ (Fig. \ref{AMOC}d) the NADW formation shuts down
and the deep AMOC is dominated by counterclockwise overturning cells.
For $\mathrm{DP^{x1}_{cl}}$ ($\mathrm{DP^{R0.1}_{cl}}$) the overturning magnitude is about $-8.5\ Sv$ ($-12.9\ Sv$) and centered at $34^\circ S$ ($30^\circ S$) and about 1500 $m$ (650 $m$) depth.
But, of course, the precise structure and magnitude of the deep counterclockwise overturning cells surely still need
several hundreds of years more in order to completely equilibrate at all depths.

\begin{figure}
\begin{center}
 \noindent\includegraphics[width=13pc]{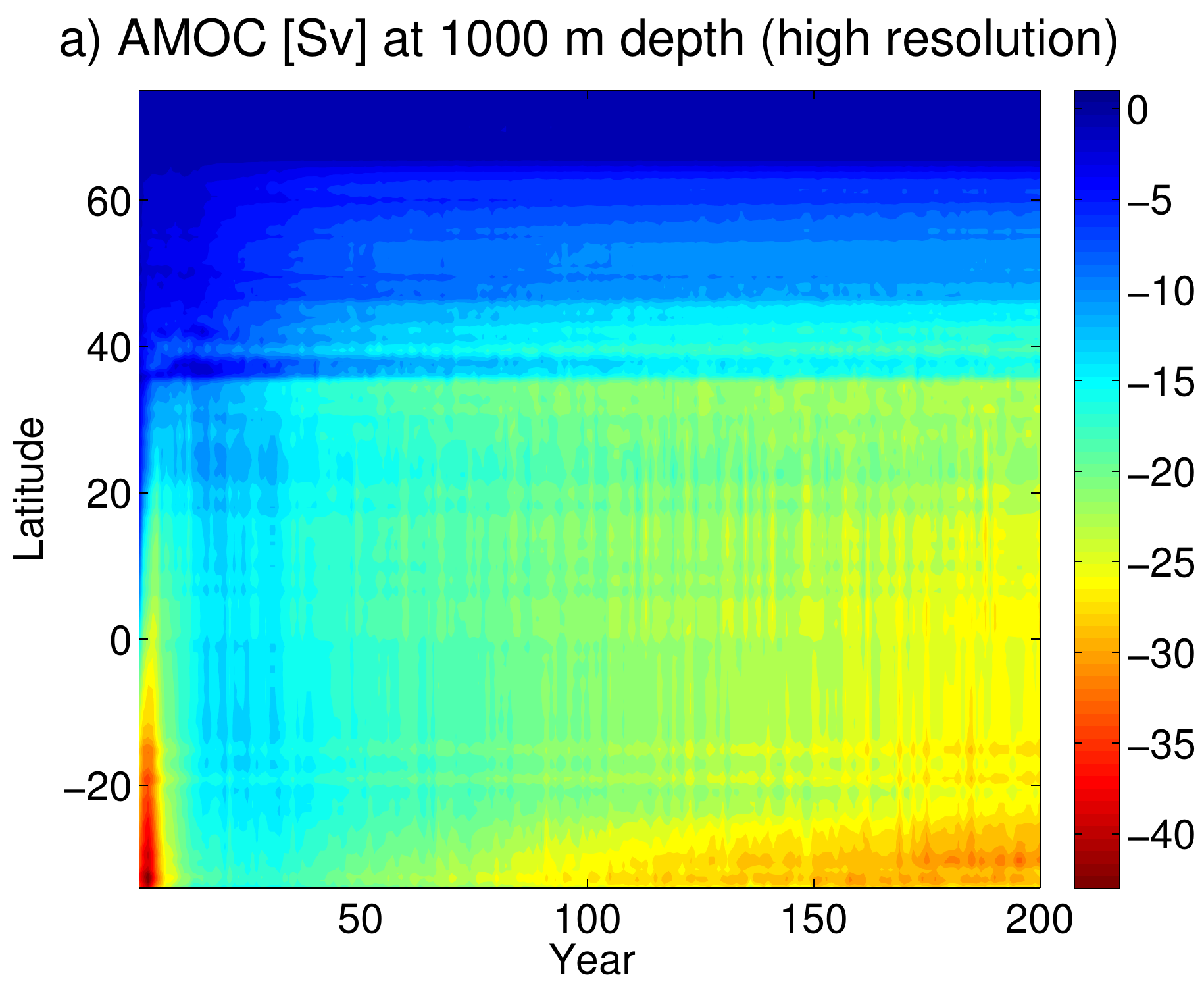}
 \noindent\includegraphics[width=13pc]{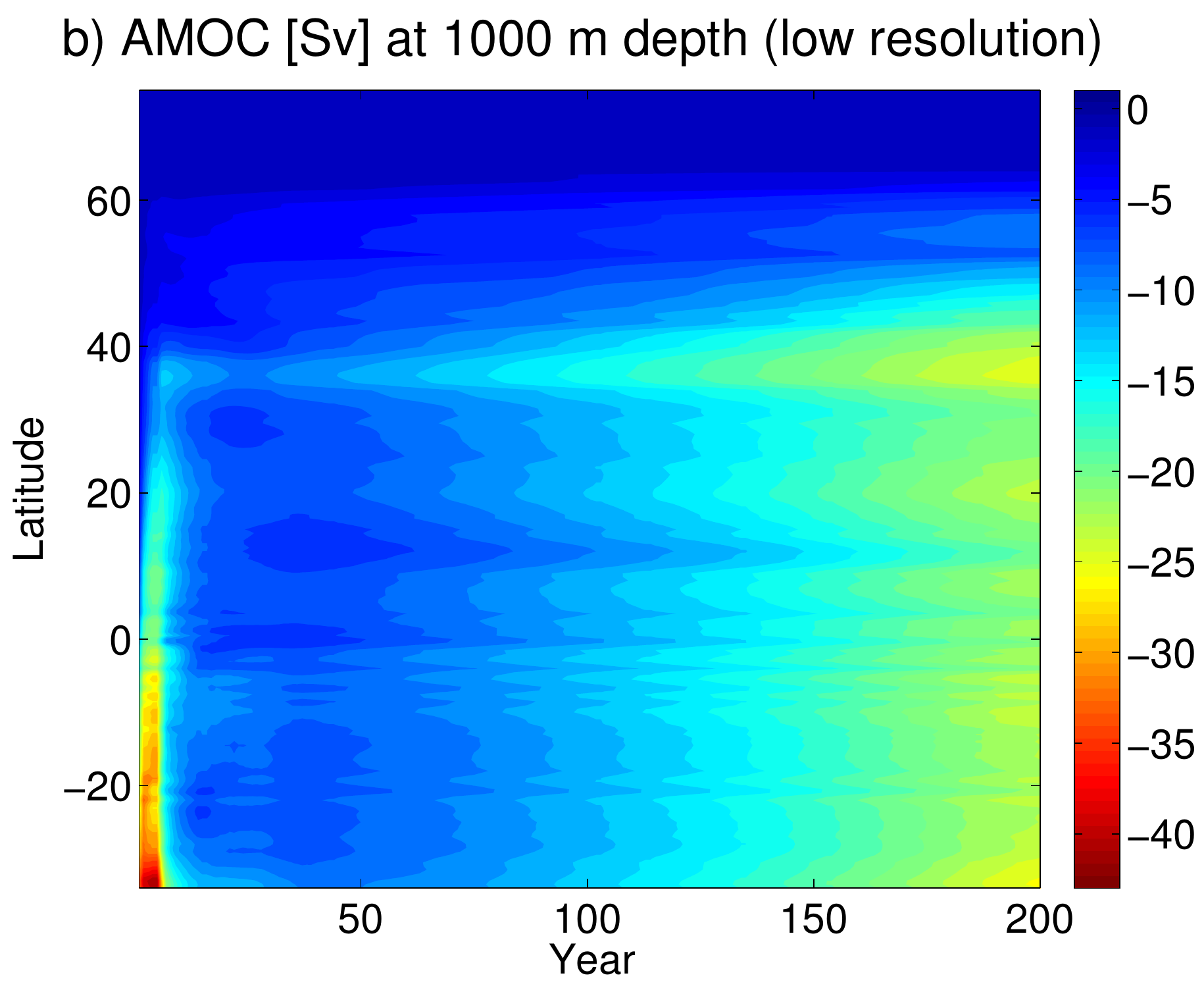}
 \noindent\includegraphics[width=13pc]{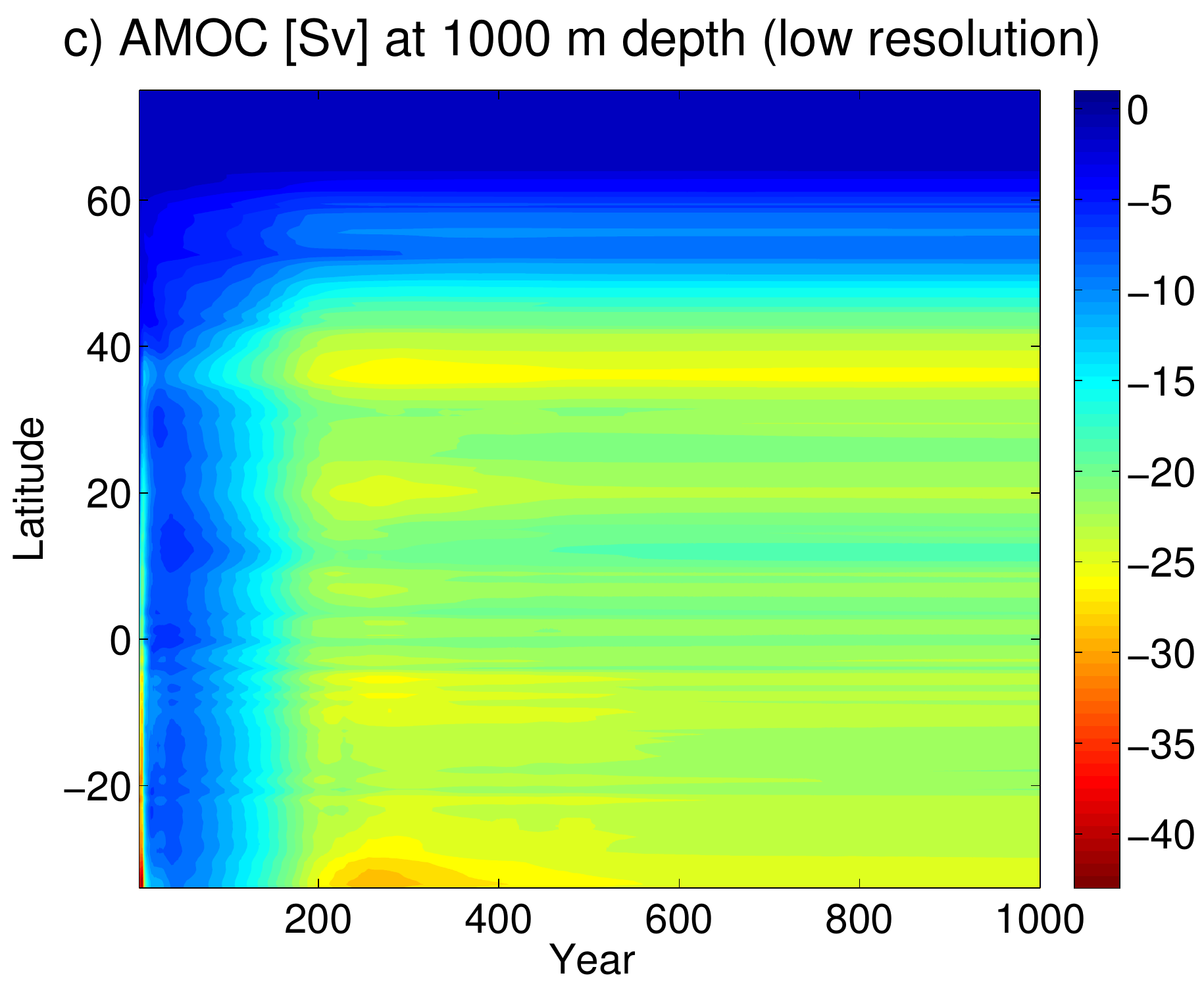}
\end{center}
\caption{Hovm\"oller diagram of the annual-mean Atlantic meridional overturning circulation [Sv] at 1000 m depth for DP closed and (a) high resolution, (b,c) low resolution (showing the first 200 years and 1000 years of the simulation).
Values are relative to the corresponding annual average for DP open at (a) reference year 76, (b,c) reference year 576.}
\label{AMOC_HOV}
\end{figure}

However, Fig. \ref{AMOC_HOV} shows Hovm\"oller diagrams of the AMOC at around $1000\ m$ depth
for both $\mathrm{DP^{R0.1}_{cl}}$ (Fig. \ref{AMOC_HOV}a) and $\mathrm{DP^{x1}_{cl}}$ (Fig. \ref{AMOC_HOV}b,c).
The Hovm\"oller diagrams indicate that the temporal evolution of the shut down of NADW formation
is very similar for both resolutions with $\mathrm{DP^{R0.1}_{cl}}$ evolving slightly faster than $\mathrm{DP^{x1}_{cl}}$.
That is, within the first 5-10 years a drastic decrease in AMOC occurs (up to $40\ Sv$ in the South Atlantic)
for both $\mathrm{DP^{R0.1}_{cl}}$ (Fig. \ref{AMOC_HOV}a) and $\mathrm{DP^{x1}_{cl}}$ (Fig. \ref{AMOC_HOV}b).
Subsequently the AMOC partially recovers for a period of 20-40 years,
whereupon it again decreases to reach its near-equilibrium values.

\section{Sea surface temperature evolution}\label{SST_sec}
\begin{figure}[tt]
 \noindent\includegraphics[width=20pc]{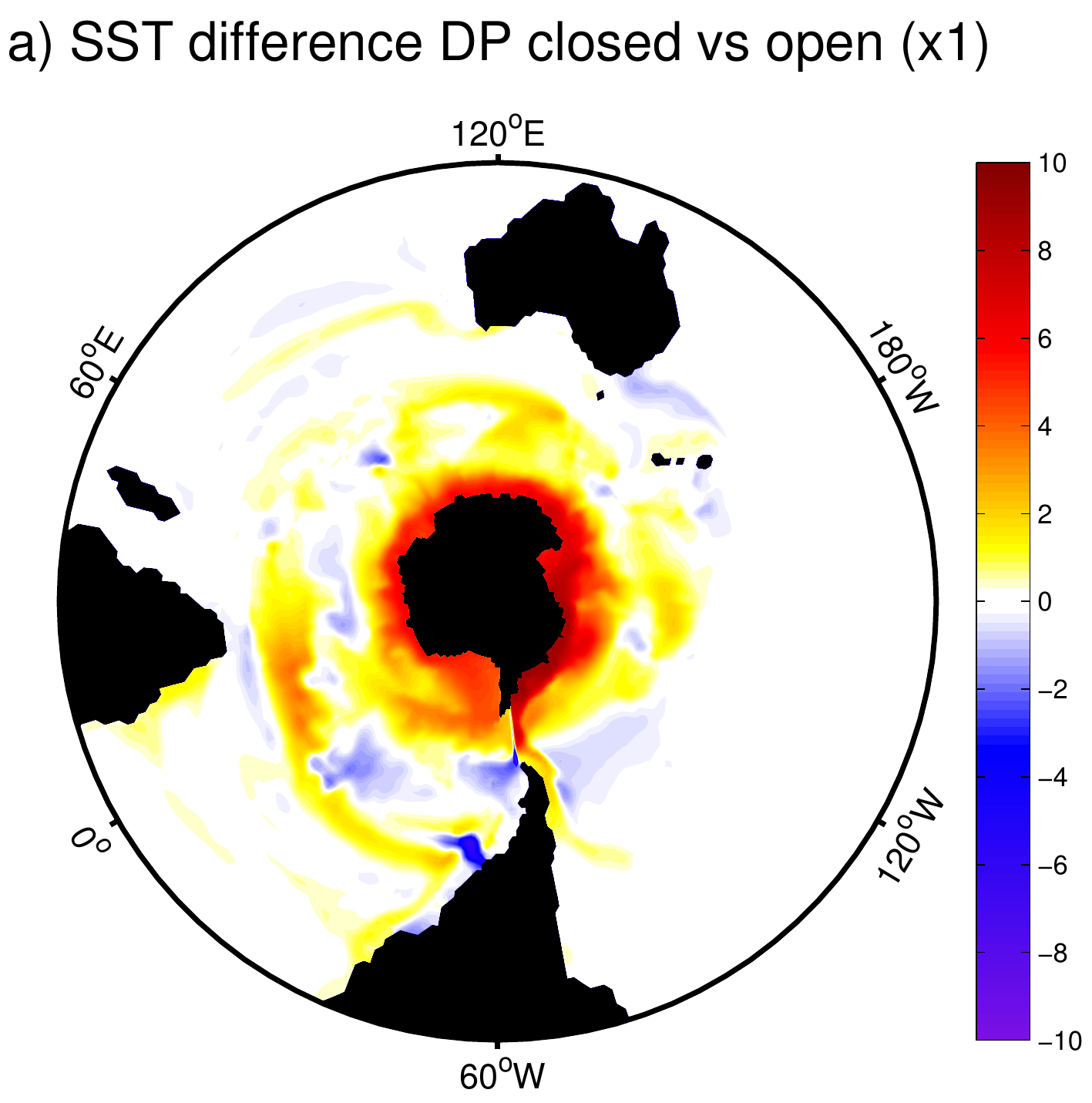}
 \noindent\includegraphics[width=20.5pc]{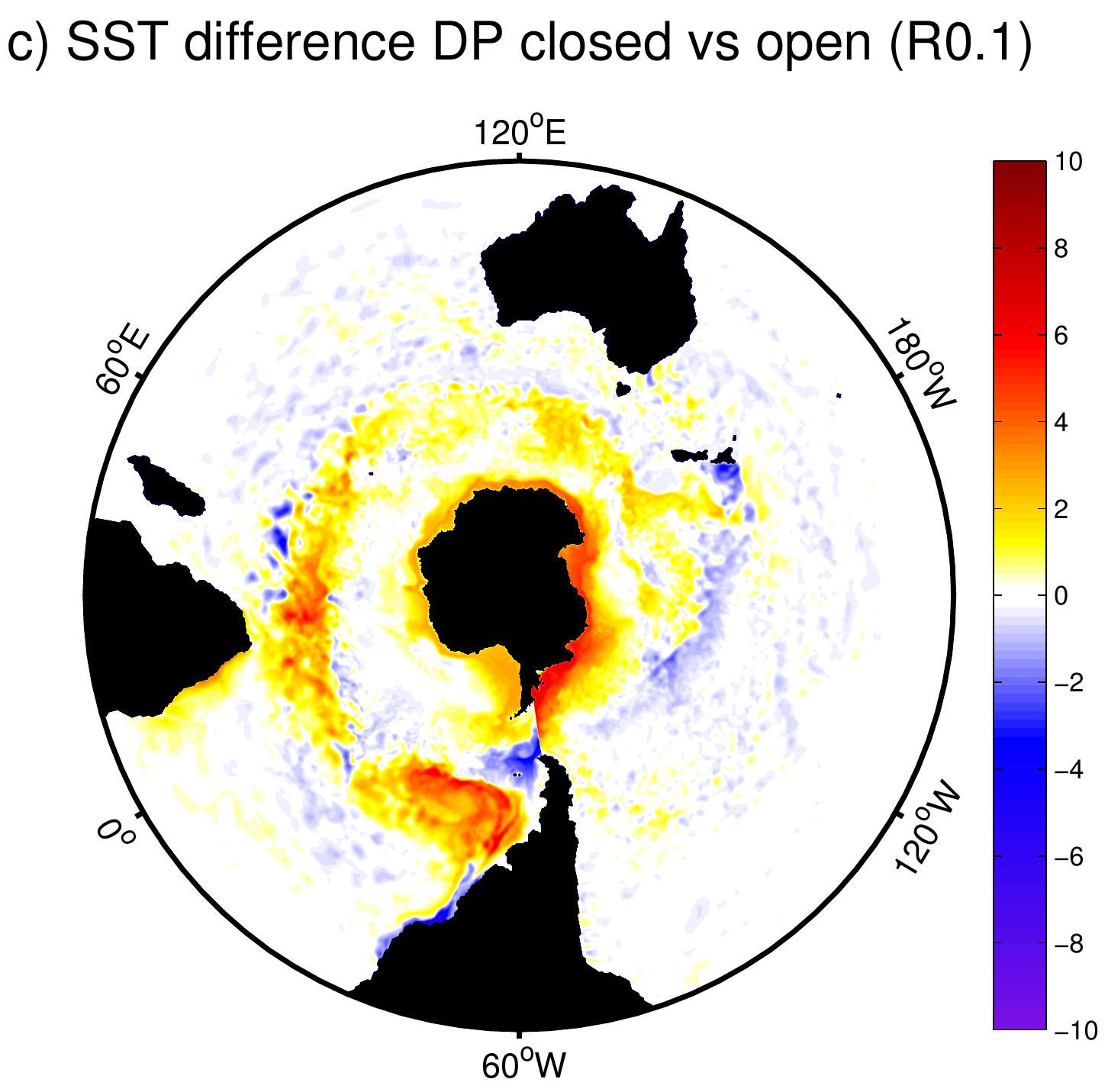}
 \noindent\includegraphics[width=20pc]{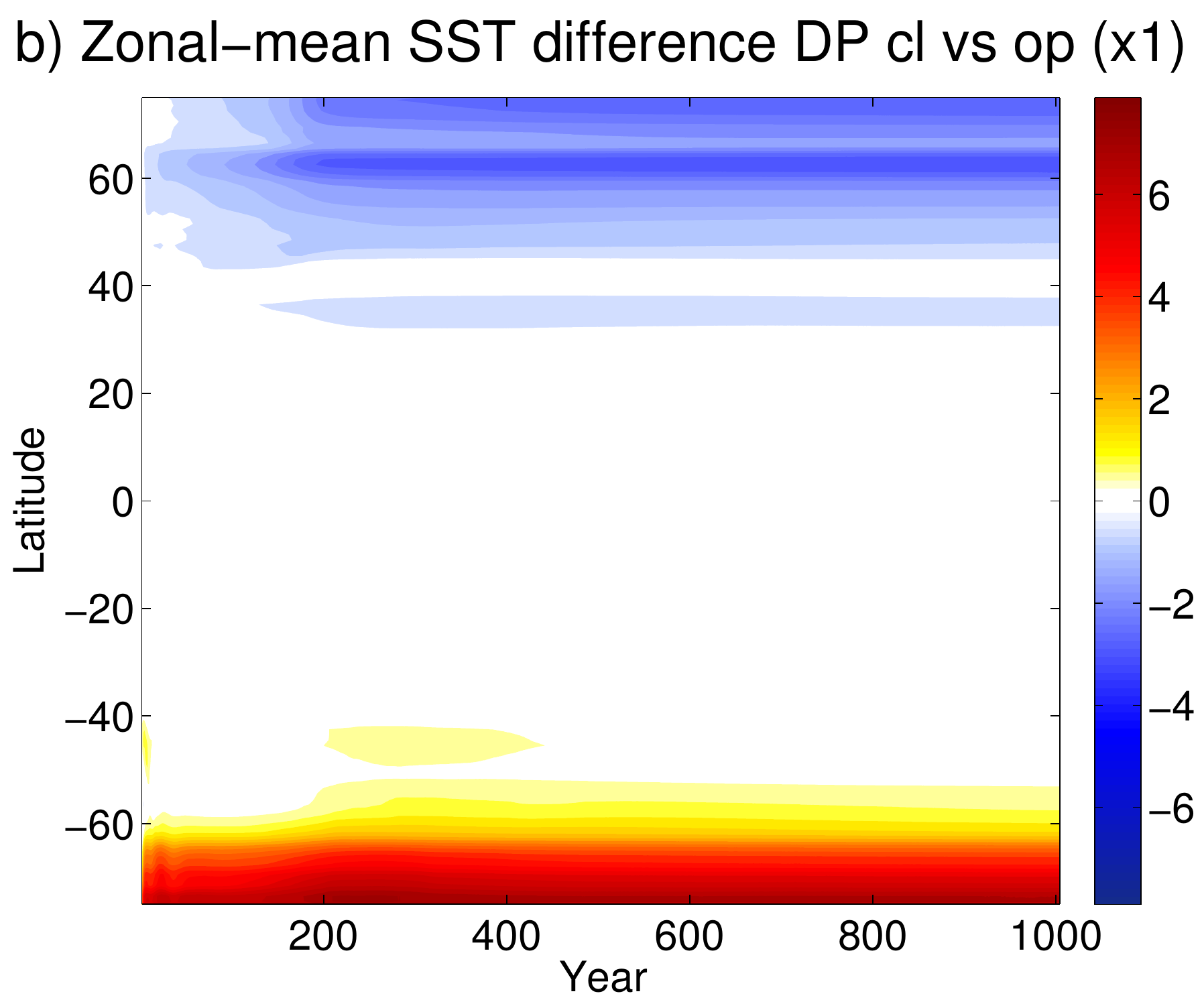}
 \noindent\includegraphics[width=20.6pc]{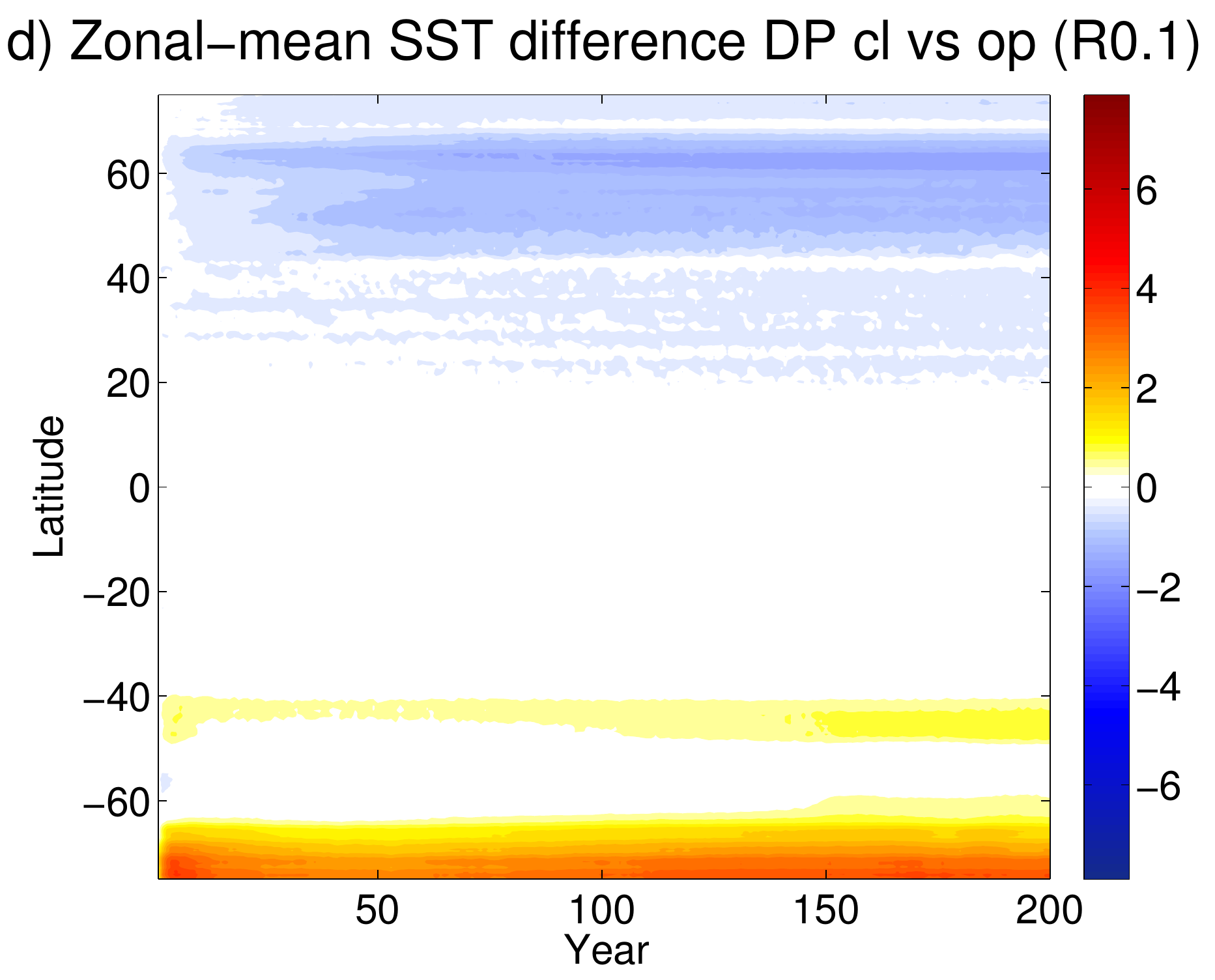}
\caption{Annual-mean SST difference [degC] between DP closed and reference year of DP open for (a,b) low resolution (reference year is 576) and (c,d) high resolution (reference year is 76).
For DP closed the years (a) 775 and (c) 275 are shown.
Panels (b,d) show Hovm\"oller diagrams of the zonal-mean SST.}
\label{SST}
\end{figure}

Figure \ref{SST} shows the changes in SST due to closing DP for both configurations $x1$ (Fig. \ref{SST}a,b) and $R0.1$ (Fig. \ref{SST}c,d).
The overall large-scale pattern of SST change is similar for both model resolutions (compare Fig. \ref{SST}a and \ref{SST}c).
The largest increase in SST is located at the Antarctic coast west of DP where subpolar gyre streamlines bend
from near-subtropical latitudes towards Antarctica.
From there the increase in SST gradually diminishes both in the zonal and the meridional directions.

The Hovm\"oller diagrams (Fig. \ref{SST}b and Fig. \ref{SST}d) indicate that for both configurations $x1$ and $R0.1$ most of this SST increase occurs within the very first years with only smaller adjustments on longer time scales.
However, the magnitude of the increase in SST around Antarctica is significantly larger for $x1$ than for $R0.1$.
For example, the increase in zonal-mean SST at $70^\circ S$ is about $5.5^\circ C$ for $\mathrm{DP^{x1}_{cl}}$ and about $2.5^\circ C$ for $\mathrm{DP^{R0.1}_{cl}}$ (with the zonal-mean SSTs of $\mathrm{DP^{x1}_{op}}$ and $\mathrm{DP^{R0.1}_{op}}$ being essentially identical).
In section \ref{MEMHT2} we demonstrate that this difference in SST change between the two model resolutions
is related to the difference in partitioning of the subpolar flow into circumpolar current and gyres (see also section \ref{barotropic_sec}).

A second local maximum in SST change occurs at a latitude band around $40^\circ S$ (see Fig. \ref{SST}a and \ref{SST}c).
For both model resolutions a hotspot of warming is located near the Agulhas retroflection region.
However, in the $R0.1$ configuration another hotspot appears in the Argentine Basin
which is much less pronounced for the x1 configuration.
This local feature is related to the bipolar mode close to Zapiola Rise \citep{Weijer:07}
and, hence, genuinely related to smaller-scale spatial variability which is not captured by
the model grid resolution of $x1$ (and the GM parameterization).

Finally, we notice that the SST in the (sub-)tropics between $35^\circ S$ and $35^\circ N$ remains largely unaffected,
whereas significant cooling occurs in the North Atlantic (see Fig. \ref{SST}b and Fig. \ref{SST}d).
The maximum cooling in zonal-mean SST is located around $60^\circ N$ (i.e. the Nordic Seas surrounding Greenland) and is about $2.0^\circ C$ ($1.3^\circ C$) for $x1$ ($R0.1$).
The temporal evolution of this SST decrease occurs on a longer time scale compared to the warming around Antarctica
and is mainly related to the shutdown of the overturning cell related to NADW formation (see also section \ref{MEMHT2}). 

\section{Meridional heat transport}\label{MHT_sec}
The ocean potential temperature $T$ is linked to the ocean flow field via the ocean heat budget.
The ocean heat budget relates the time tendency of  $T$ (i.e. ocean heat content) to both
the divergence of heat transport by the ocean flow field and the surface heat fluxes \citep{Griffies:15,Yang:15}.
Within the longitude and depth integrated heat budget the ocean flow field predominantly enters via 
the advective meridional heat transport,
\begin{equation}
MHT\equiv C_p\rho_0\iint vT\ dzdx\ ,
\end{equation}
where $v$ represents the meridional velocity (also including the GM part for $x1$),
$\rho_0$ the reference density, and $C_p$ the ocean heat capacity.
That is, if there is no regional storage 
of heat within the ocean 
and contributions to the meridional heat transport by subgrid scale processes (e.g. diffusion)
are small,
then the air-sea heat fluxes are balanced by the meridional divergence of $MHT$ at each latitude.

\subsection{Global and basin-wide depth integrated heat advection}\label{MEMHT0}
\begin{figure}
 \noindent\includegraphics[width=20pc]{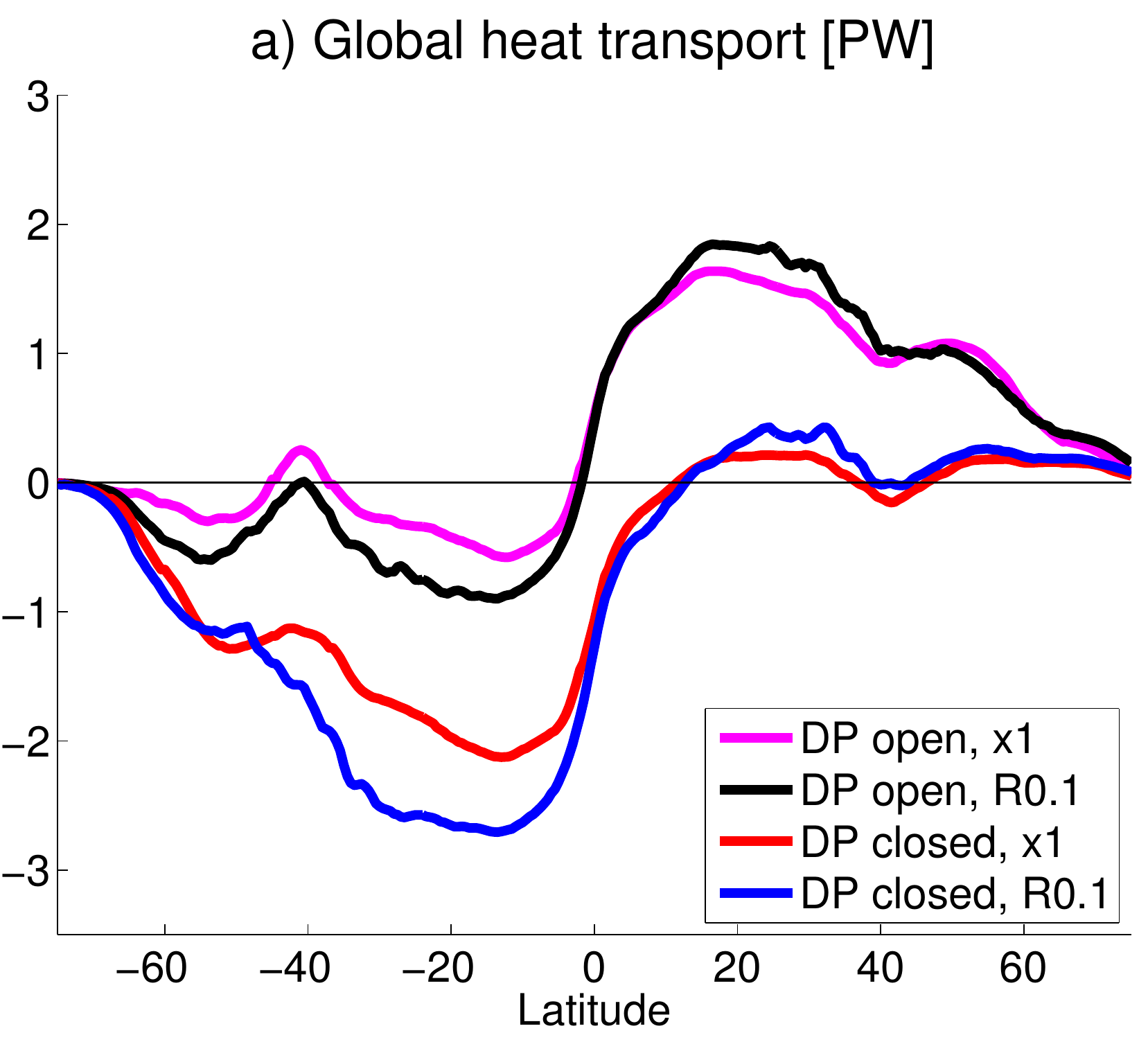}
 \noindent\includegraphics[width=20pc]{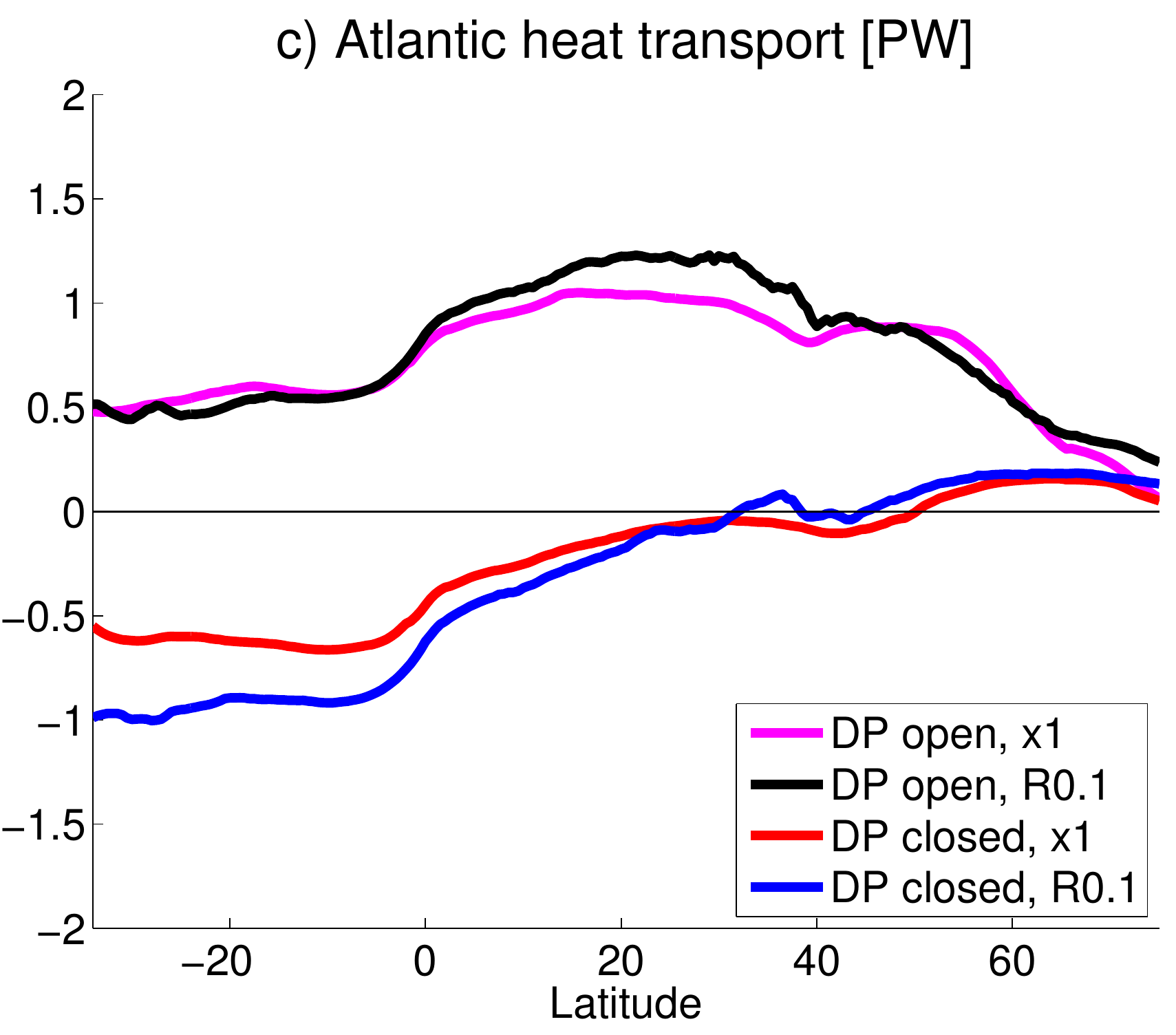}
 \noindent\includegraphics[width=20pc]{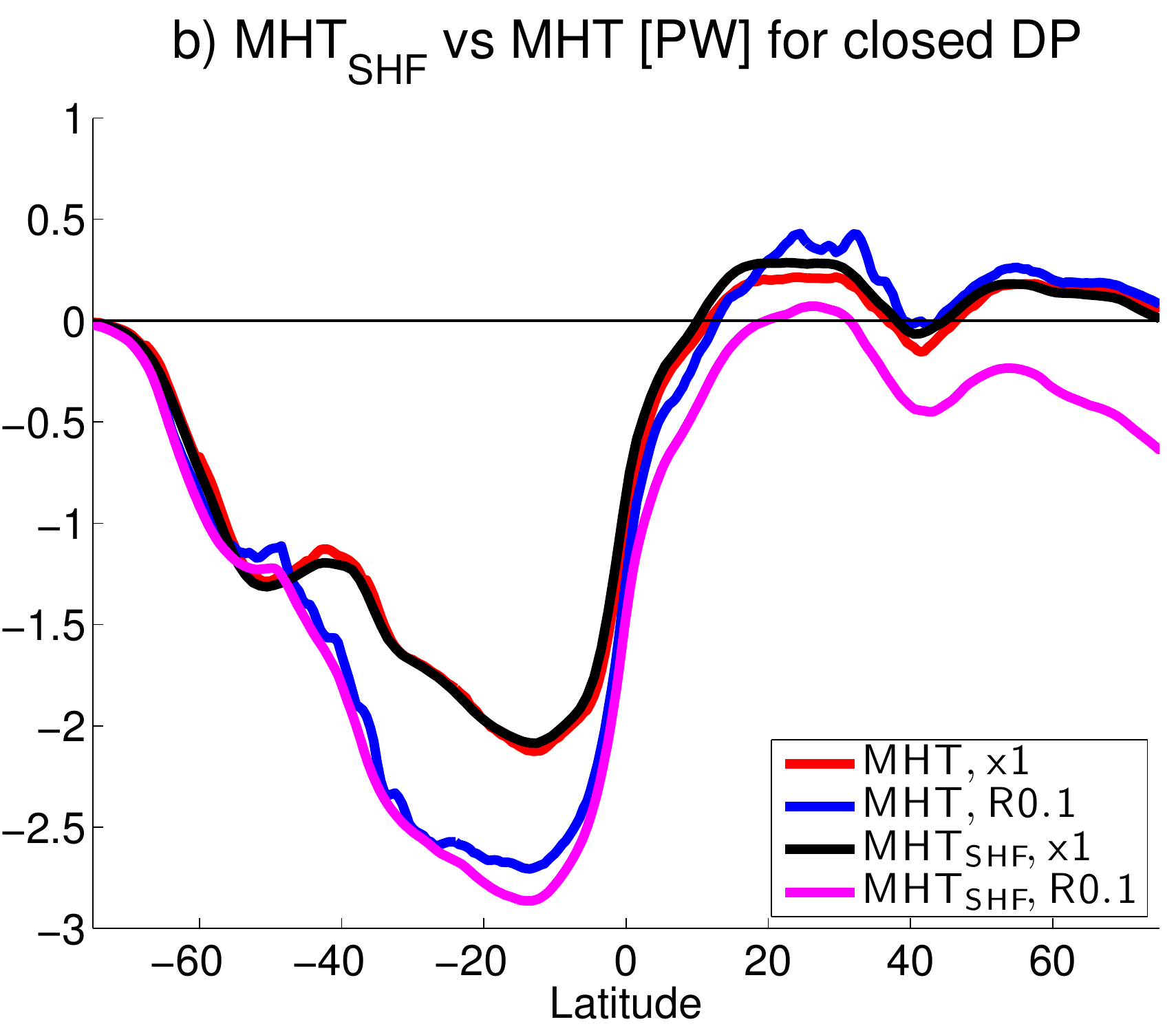}
 \noindent\includegraphics[width=20pc]{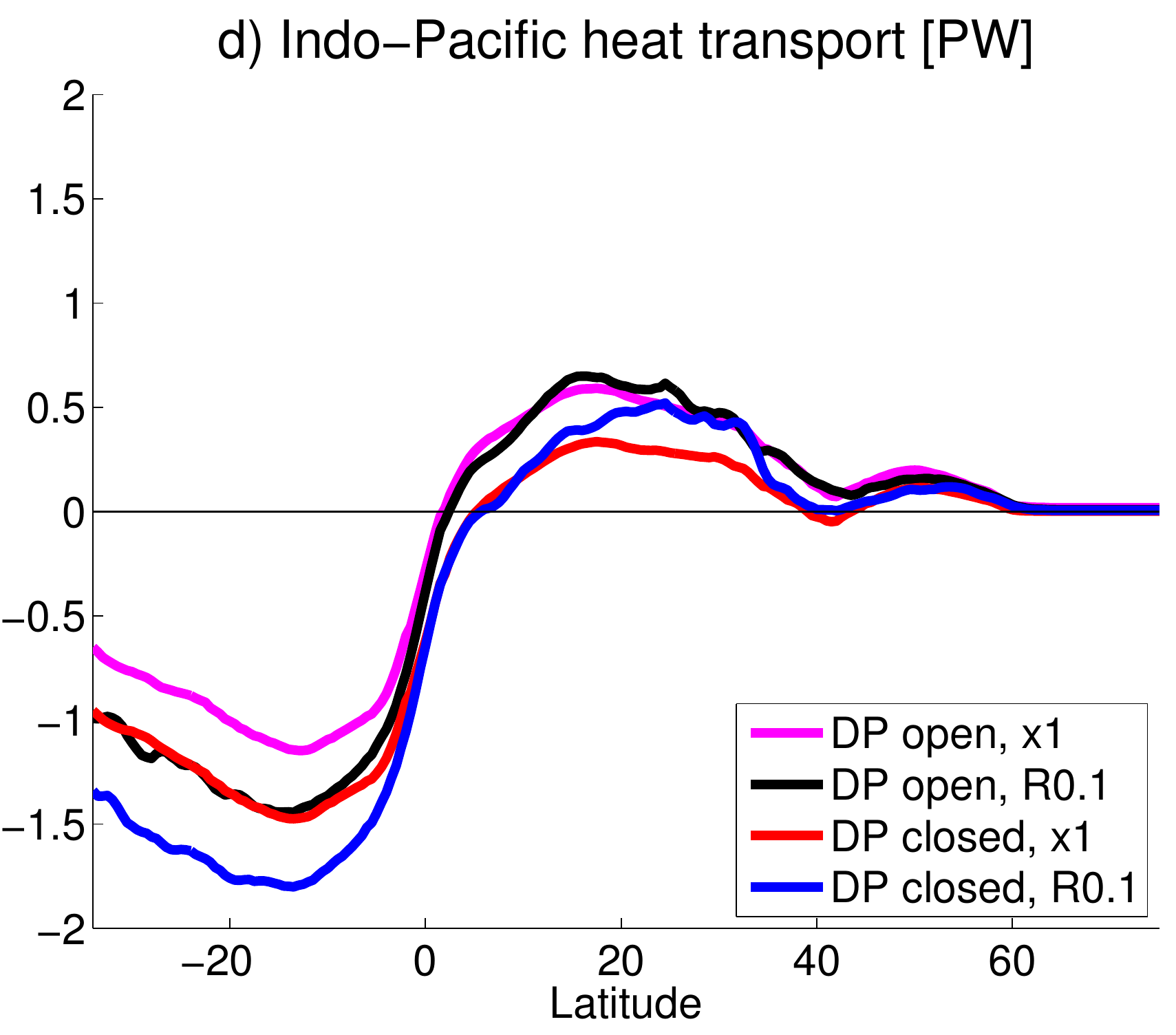}
\caption{The (a) global, (c) Atlantic-Arctic, and (d) Indo-Pacific annual-mean advective meridional heat transports [PW] for all model simulations are shown. For the control simulations the reference years 76 (for $\mathrm{DP^{R0.1}_{op}}$)
and 576 (for $\mathrm{DP^{x1}_{op}}$) are presented (the curves remain essentially unchanged for later years). 
For the closed DP simulations the years 275 (for $\mathrm{DP^{R0.1}_{cl}}$) and 1575 (for $\mathrm{DP^{x1}_{cl}}$)
are shown (see additionally note \ref{fn2}).
Also (b) the zonally and meridionally integrated (from south to north) surface heat flux ($MHT_{SHF}$) is drawn for the closed DP simulations.
For comparison the corresponding global advective meridional heat transport is shown again (same as in panel a).}
\label{MHT_SHF1}
\end{figure}

Figure \ref{MHT_SHF1} displays the annual-mean $MHT$ for the global ocean (Fig. \ref{MHT_SHF1}a) as well as its decomposition into Atlantic-Arctic (AA, Fig. \ref{MHT_SHF1}c) and Indo-Pacific (IP, Fig. \ref{MHT_SHF1}d) components.
For the control simulations $\mathrm{DP^{R0.1}_{op}}$ (black curve) and $\mathrm{DP^{x1}_{op}}$ (magenta curve)
the overall behavior is very similar exhibiting the present-day structure of a hemispherically antisymmetric distribution in the IP (i.e. $MHT$ from the equator to the poles which is slightly larger in the Southern Hemisphere (SH) due to the addition of the Indian Ocean)
but northward $MHT$ in the entire AA (due to the overturning circulation related to NADW formation).
Consequently, the global $MHT$ from the equator to the poles is significantly larger in the Northern Hemisphere (NH)
than in the SH with the local minimum around $40^\circ S$ (related to the boundary between subtropical gyres and ACC) even showing small equatorward transport.
The main 
difference between $\mathrm{DP^{R0.1}_{op}}$ and $\mathrm{DP^{x1}_{op}}$ is that the magnitude of $MHT$ is slightly larger in $\mathrm{DP^{R0.1}_{op}}$ than in $\mathrm{DP^{x1}_{op}}$,
as it is found in other recent studies comparing the $MHT$ of low-resolution and high-resolution present-day climate
model simulations \citep{Bryan:14,Griffies:15}.
More precisely, for $\mathrm{DP^{R0.1}_{op}}$  ($\mathrm{DP^{x1}_{op}}$) the maximum global/AA/IP $MHT$ is 1.85/1.23/0.65 PW (1.64/1.05/0.59 PW),
and the minimum global/IP $MHT$ 
is -0.9/-1.44 PW (-0.58/-1.15 PW).
That is, in the NH the smaller $MHT$ of $\mathrm{DP^{x1}_{op}}$ is mostly located in the Atlantic,
whereas in (sub-)tropics of the SH smaller $MHT$ is mainly present in the IP.
However, most crucial for this study 
is that in the SO (i.e. south of $35^\circ S$) the $MHT$ of $\mathrm{DP^{x1}_{op}}$ is smaller as well (Fig. \ref{MHT_SHF1}a).

Figure \ref{MHT_SHF1} also shows the corresponding annual-mean $MHT$ for both $\mathrm{DP^{R0.1}_{cl}}$ (blue curve) and $\mathrm{DP^{x1}_{cl}}$ (red curve) with again very similar overall behavior for both resolutions.
The magnitude of the global $MHT$ (Fig. \ref{MHT_SHF1}a) is drastically reduced in the NH and drastically increased in the SH leading to the reversed distribution of $\mathrm{DP_{op}}$.
That is, the global $MHT$ from the equator to the poles is significantly larger in the SH
than in the NH with the local minimum around $40^\circ N$ (related to the boundary between subtropical and subpolar gyres) even showing small equatorward transport.
The maximum change is of about $2$ PW and is mostly located in the AA (maximum change of about $1.5$ PW, Fig. \ref{MHT_SHF1}c)
where the $MHT$ turns into southward $MHT$ within the entire (sub-)tropics ($35^\circ S<y<35^\circ N$).
The IP (Fig. \ref{MHT_SHF1}d) keeps a hemispherically antisymmetric distribution ($MHT$ from the equator to the poles) but the magnitude of $MHT$ is increased in the SH (up to $0.5$ PW) and decreased in the NH.
Again the main difference between the two configurations is that the maximum values in $\mathrm{DP^{R0.1}_{cl}}$ are generally larger compared\footnote{\label{fn2}
We note that the precise quantitative differences between $\mathrm{DP^{R0.1}_{cl}}$ and $\mathrm{DP^{x1}_{cl}}$
shown in Fig. \ref{MHT_SHF1} are still subject to the remaining small adjustment processes active in $\mathrm{DP^{R0.1}_{cl}}$, and are expected to partly reduce with longer simulation time of $\mathrm{DP^{R0.1}_{cl}}$.
More precisely, the global $MHT$s of $\mathrm{DP^{x1}_{cl}}$ at years $200$ and $1000$ are very similar,
whereas at year 300  the values in the Southern Hemisphere (sub-)tropics are closer to the shown values of $\mathrm{DP^{R0.1}_{cl}}$ at year 200.
This is related to the fact that $\mathrm{DP^{R0.1}_{cl}}$ has a slightly shorter adjustment time scale than $\mathrm{DP^{x1}_{cl}}$ (as can be seen e.g. in the Hovm\"oller diagrams of Fig. \ref{AMOC_HOV})
such that e.g. year 200 of $\mathrm{DP^{R0.1}_{cl}}$ is more similar to year 300 of $\mathrm{DP^{x1}_{cl}}$.}
to $\mathrm{DP^{x1}_{cl}}$.
However, we observe that within the SO the difference in $MHT$ between the configurations is significantly reduced
with nearly identical values south of $50^\circ S$ (Fig. \ref{MHT_SHF1}a).

Regarding the adjustment processes that are still active in the DP closed simulations
we note that the air-sea heat fluxes can be integrated meridionally 
to obtain an estimate of $MHT$ necessary to balance the loss or gain of heat through air-sea exchange.
Observational estimates of the meridional heat transport are typically based on this indirect calculation via air-sea heat fluxes \citep{Macdonald:13}.
Within numerical ocean models the comparison of direct and indirect meridional heat transport computations
gives an indication of the equilibration of the ocean state and whether fundamental physical constraints like mass and energy conservation are satisfied \citep{Yang:15}.
Figure \ref{MHT_SHF1}b shows both the zonally and meridionally integrated (from south to north) surface heat flux (indicated by $MHT_{SHF}$) and the global $MHT$ (same as in \ref{MHT_SHF1}a) for both $\mathrm{DP^{R0.1}_{cl}}$ and $\mathrm{DP^{x1}_{cl}}$.
For $\mathrm{DP^{x1}_{cl}}$ the $MHT$ and $MHT_{SHF}$
nearly overlap which demonstrates that the ocean heat content is largely equilibrated at every latitude.
For $\mathrm{DP^{R0.1}_{cl}}$ on the other hand,
the ocean heat content is still slightly adjusting within subtropics of the SH as well as
more substantially within the NH.
Nevertheless, south of $50^\circ S$ the $MHT$ and $MHT_{SHF}$ almost perfectly overlap.
This indicates that the SO is the region where the ocean heat content reaches a near-equilibrium state first,
in correspondence with the Hovm\"oller diagrams of SST (Fig. \ref{SST}b,d).

However, crucial questions remain:
What are the respective roles of the mean flow and the eddy field in the $MHT$ for the simulations $\mathrm{DP_{op}}$ and $\mathrm{DP_{cl}}$?
What are the respective contributions of the barotropic circulation (section \ref{barotropic_sec}) and the overturning circulation (section \ref{overturning_sec}) to the corresponding (changes in) $MHT$?
Do the two configurations $R0.1$ and $x1$ behave differently in these respects?
In previous paleoceanographic gateway studies
these attribution questions (i.e. the partitioning of $MHT$ into different dynamical components)
were not or only loosely addressed.
But in order to understand changes in $MHT$ it is mandatory to determine what physical processes control the $MHT$ \citep{Ferrari:11}. 
In order to relate the (changes in) $MHT$ more precisely to
individual contributions of different physical processes
we perform both temporal and spatial decompositions of the global $MHT$ in the following two sections.

\subsection{Temporal decomposition of the depth integrated heat advection}\label{MEMHT1}
We decompose the global $MHT$ into a time-mean component,
\begin{equation}
\overline{MHT}\equiv C_p\rho_0\iint\overline{v}^t\overline{T}^t\ dzdx\ ,
\end{equation}
and a transient eddy component (also including the GM part for $x1$),
\begin{equation}
MHT'\equiv C_p\rho_0\iint\overline{vT}^t-\overline{v}^t\overline{T}^t\ dzdx\ ,
\end{equation}
based on the temporal mean $\overline{\cdot}^t$.
We note that the variability within the non-eddying configuration ($x1$) is essentially given by the
seasonal cycle,
whereas within the eddying configuration ($R0.1$) also internal variability on inter-annual to decadal time scales is present \citep{LeBars:15}.
Consequently, the eddy-mean decomposition of the eddying configuration partly depends on the chosen length of the time averaging period.
Different averaging intervals affect mostly $MHT'$ due to its relatively small magnitudes,
whereas $MHT$ and $\overline{MHT}$ remain 
largely unchanged.
For that reason we discuss $MHT'$ with respect to both $1$-year and $20$-year average periods in the following.

Figure \ref{MHT_SHF3}a shows the annual-mean $\overline{MHT}$ for all simulations.
In each case $\overline{MHT}$ follows largely the corresponding $MHT$ (compare with Fig. \ref{MHT_SHF1}a)
such that the overall similarity between $\overline{MHT}$ and $MHT$
typically found for $\mathrm{DP_{op}}$ \citep{Bryan:14,Griffies:15,Yang:15} equally holds for $\mathrm{DP_{cl}}$.
In particular, similar to $MHT$ we find that the magnitudes of $\overline{MHT}$ are generally larger for
the eddying configuration but that south of $50^\circ S$ both $\mathrm{DP^{R0.1}_{cl}}$ and $\mathrm{DP^{x1}_{cl}}$
show nearly identical values (in contrast to $\mathrm{DP_{op}}$).

\begin{figure}
 \noindent\includegraphics[width=20pc]{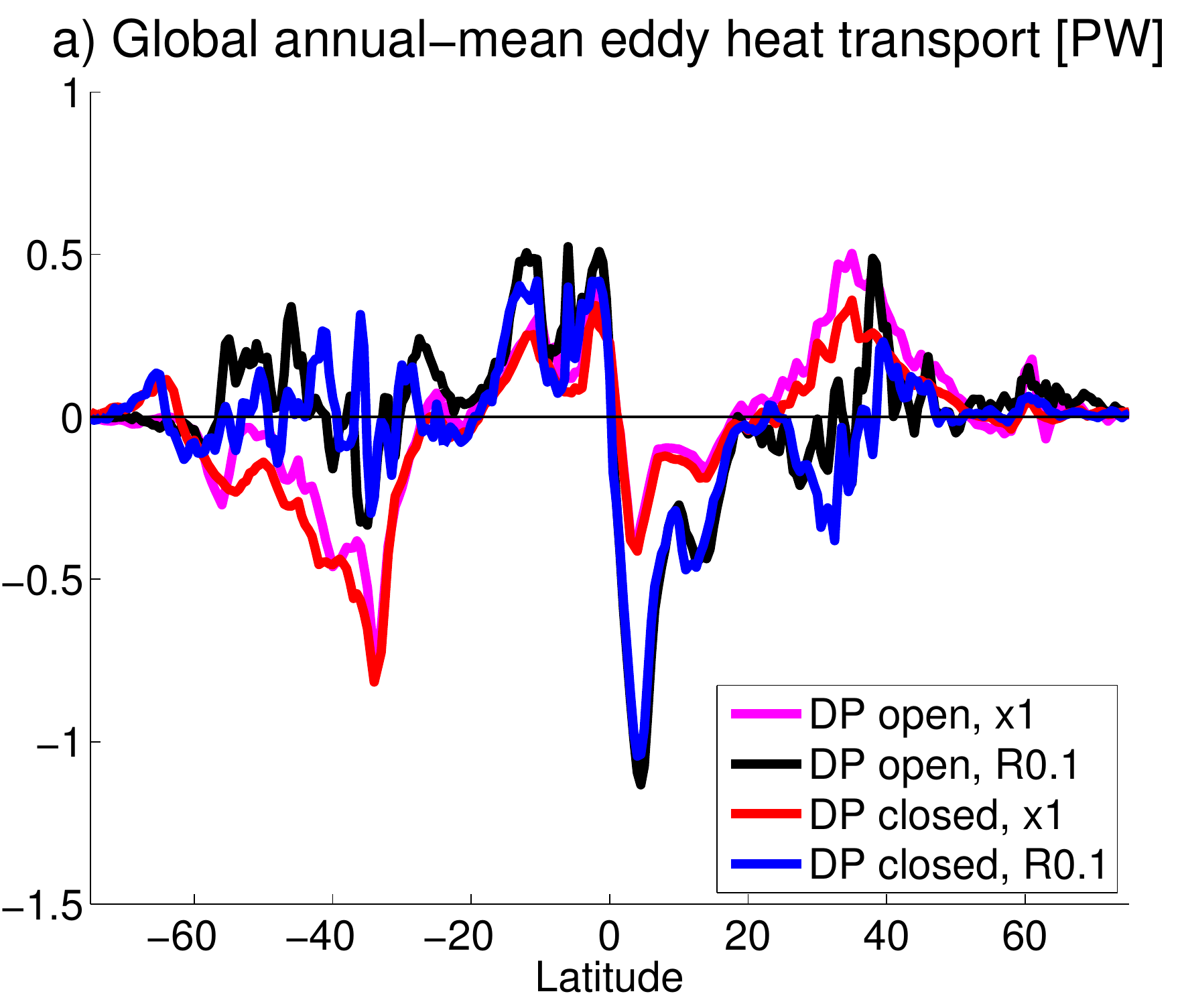}
 \noindent\includegraphics[width=20pc]{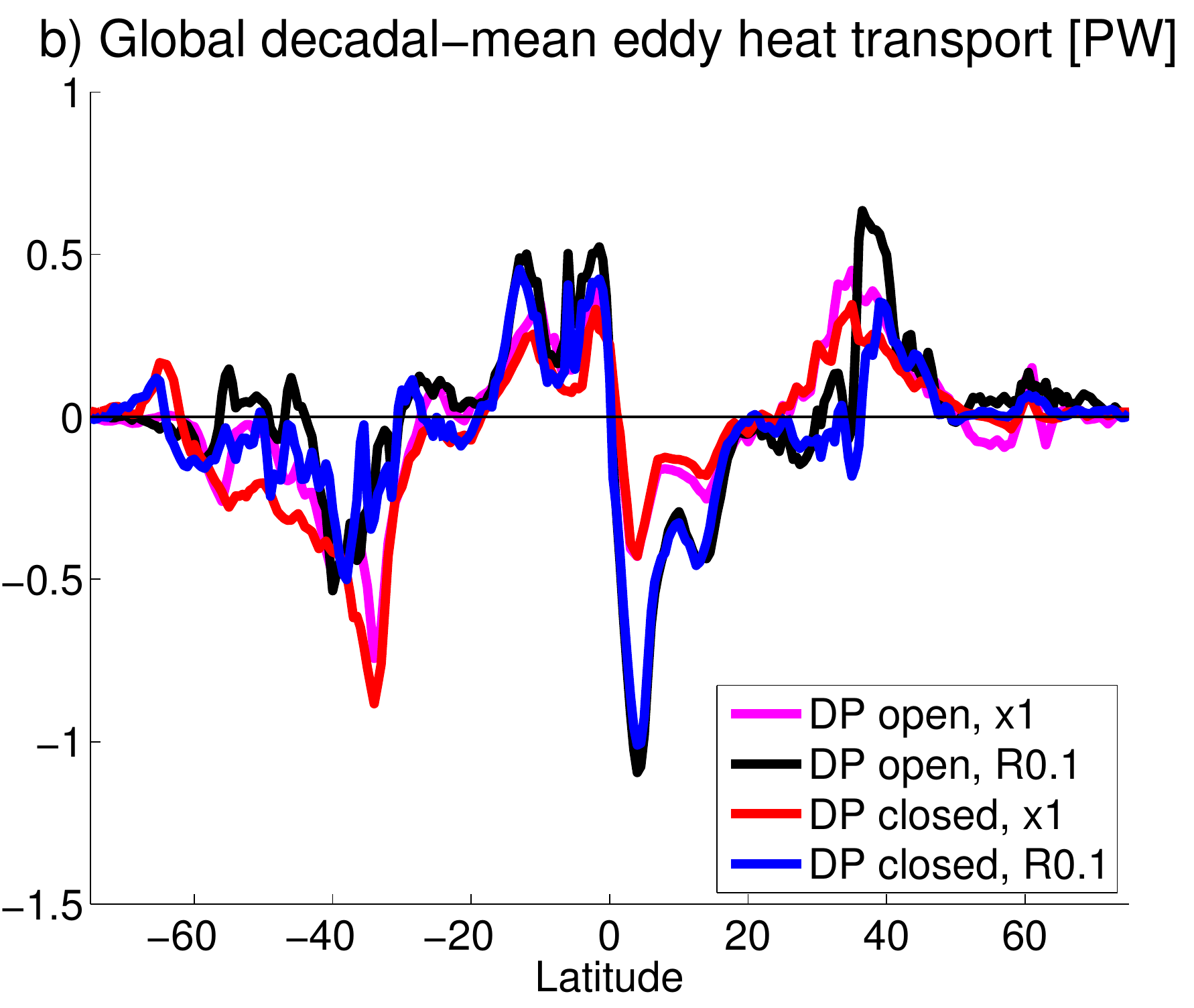}
\caption{The eddy component of the global advective meridional heat transport [PW] is shown for all simulations
and two averaging intervals.
In (a) annual averages are used for the same years as in Fig. \ref{MHT_SHF1}.
In (b) 20-year averages are performed of years 76-95 (for $\mathrm{DP^{R0.1}_{op}}$),
576-595 (for $\mathrm{DP^{x1}_{op}}$), 256-275 (for $\mathrm{DP^{R0.1}_{cl}}$), and 756-775 (for $\mathrm{DP^{x1}_{cl}}$).\\
\ \\\ \\\ \\\ \\
}
\label{MHT_SHF2}
\end{figure}

The difference between $MHT$ and $\overline{MHT}$, i.e. $MHT'$,
is shown in Fig. \ref{MHT_SHF2} for both annual (Fig. \ref{MHT_SHF2}a) and multidecadal (Fig. \ref{MHT_SHF2}b) averaging periods.
Focussing on $\mathrm{DP^{x1}_{op}}$ (magenta line, essentially unaffected by the averaging period)
we find the structure of $MHT'$ typically described
in previous studies using decadal averaging periods \citep{Bryan:96,Jayne:02,Volkov:08,Bryan:14,Griffies:15}.
Significant magnitudes of $MHT'$ (i.e. in the order of $\overline{MHT}$)
are found in three main regions:
On the one hand, a strong convergence of $MHT'$ opposing $\overline{MHT}$ is found in the tropics and
related to tropical instability waves;
on the other hand, around $35^\circ S$ and $35^\circ N$ peaks of poleward $MHT'$ are present
and related to western boundary currents (e.g. the Kuroshio Current or the Brazil-Malvinas Confluence regions)
as well as their extensions into the ocean interior (i.e. eddy-induced heat transport across gyre boundaries),
the Agulhas Retroflection and the ACC.

The $MHT'$ of $\mathrm{DP^{x1}_{cl}}$ exhibits exactly the same structure (compare red and magenta lines)
with slightly smaller values in the tropics and around $35^\circ N$ and slightly larger values around $35^\circ S$.
This largely insensitive behavior may indicate that the $MHT'$ is dominated by wind-driven instabilities
which are largely determined by the prescribed wind stress.
One exception occurs south of $60^\circ S$
where the nearly vanishing southward $MHT'$ of $\mathrm{DP^{x1}_{op}}$
turns into a substantial northward $MHT'$ for $\mathrm{DP^{x1}_{cl}}$ (red line).
This eddy transport is induced by the intensification
of the subpolar gyre Antarctic boundary current
due to the integration of the former ACC into the subpolar gyre system.

The $MHT'$s of both $\mathrm{DP^{R0.1}_{op}}$ (black line) and $\mathrm{DP^{R0.1}_{cl}}$ (blue line) related to multidecadal averages (Fig. \ref{MHT_SHF2}b) show a similar structure as the corresponding $MHT'$ of the non-eddying simulations.
In particular, for a closed DP the $MHT'$ decreases in the tropics and NH,
whereas it slightly increases around $40^\circ S$ and a substantial northward $MHT'$ emerges near Antarctica.
Compared to $\mathrm{DP^{x1}}$ the magnitudes are larger in the tropics
but smaller south of  $20^\circ S$.
North of $20^\circ N$ the magnitudes are also smaller except for the strong peak around $40^\circ N$.
These quantitative differences are largely related to the different pathways
of western boundary current extensions for different resolutions \citep{Griffies:15}.

Finally, Fig. \ref{MHT_SHF2}a also shows the $MHT'$s of both $\mathrm{DP^{R0.1}_{op}}$ (black line) and $\mathrm{DP^{R0.1}_{cl}}$ (blue line) for annual averages.
Within the tropics and south of $60^\circ S$ (as well as north of $50^\circ N$) the $MHT'$
is largely unaffected by the time averaging period.
However, in between these regions significant differences appear
with $MHT'$ having smaller magnitudes and partly turning from poleward to equatorward.
These differences are related to inter-annual variability
and show that the $MHT'$ of the non-eddying simulations (based on the GM parameterization)
gives an adequate representation of the $MHT'$ of the eddying simulations generally only in a decadal-mean sense (Fig. \ref{MHT_SHF2}b).
Nevertheless, the low-resolution $MHT'$ is able to capture the qualitative changes south of $60^\circ S$ for both averaging intervals in our simulations.

In summary, closing DP induces only relatively small quantitative changes in the $MHT'$.
The only qualitative change is the emergence of northward transport south of $60^\circ S$
which is well captured by the non-eddying simulation too.
Consequently, the changes in $MHT$ (Fig. \ref{MHT_SHF1}a) due to a closed DP
are dominated by the changes in $\overline{MHT}$ (Fig. \ref{MHT_SHF3}a).
For that reason we decompose $\overline{MHT}$ in the following section
in order to relate the circulation changes described in section \ref{flowfield} to the changes in $MHT$.


\subsection{Spatial decomposition of the time-mean depth integrated heat advection}\label{MEMHT2}
We decompose the global $\overline{MHT}$ based on the global zonal mean $\overline{\cdot}^x$
into a component related to time-mean meridional overturning circulations,
\begin{equation}\label{MHTO}
\overline{MHT}_O\equiv C_p\rho_0\iint\overline{v}^{tx}\overline{T}^t\ dzdx\ ,
\end{equation}
and a component related to time-mean horizontal 
circulations,
\begin{equation}\label{MHTH}
\overline{MHT}_H\equiv C_p\rho_0\iint(\overline{v}^{t}-\overline{v}^{tx})\overline{T}^t\ dzdx\ ,
\end{equation}
as introduced by \cite{Hall:82} (see also \cite{Bryden:01,Macdonald:13}) and applied to the SO e.g. by \cite{Treguier:07,Volkov:10}.

We note that this decomposition does not entirely disentangle
the respective heat transport contributions of wind-driven gyre circulations and deep overturning circulations.
For example, $\overline{MHT}_O$ includes both warm shallow wind-driven overturning cells
as well as cold deep overturning cells.
Approaches to distinguish the heat transport related to individual meridional overturning cells via a so-called heatfunction
were introduced by \cite{Boccaletti:05,Greatbatch:07,Ferrari:11} and applied e.g. by \cite{Yang:15}.
However, these approaches do not separate the contributions by horizontal 
circulations which are captured by $\overline{MHT}_H$ and mainly related to wind-driven circulations.
Since we are interested into the changes in $MHT$ due to the reorganization
of the wind-driven horizontal circulation in the SO
(integration of the ACC into the subpolar gyre system, section \ref{barotropic_sec})
the decomposition by Eq. (\ref{MHTO})-(\ref{MHTH}) appears more appropriate for this study.
Nevertheless, we encourage the application of other decompositions of $MHT$ (e.g. via the heatfunction approach)
in future paleoceanographic studies\footnote{
In order to be able to distinguish within the heatfunction framework of \cite{Ferrari:11}
between heat transport by overturning circulations and heat transport by horizontal circulations
one would have to extend their framework
and perform a standing eddy decomposition of the isothermal streamfunction
as discussed in \cite{Viebahn:12}.
The heat transport by horizontal circulations would be captured by the standing eddy streamfunction
with significant values mainly in the upper layers related to wind-driven gyres (except for the ACC belt region).
Decomposing the heatfunction might also shed light on the ``mixed mode'', that is,
the heat transport that can not be uniquely attributed to either gyre overturning or deep overturning.
However, such an analysis is beyond the scope of this study.
}.

\begin{figure}
 \noindent\includegraphics[width=13pc]{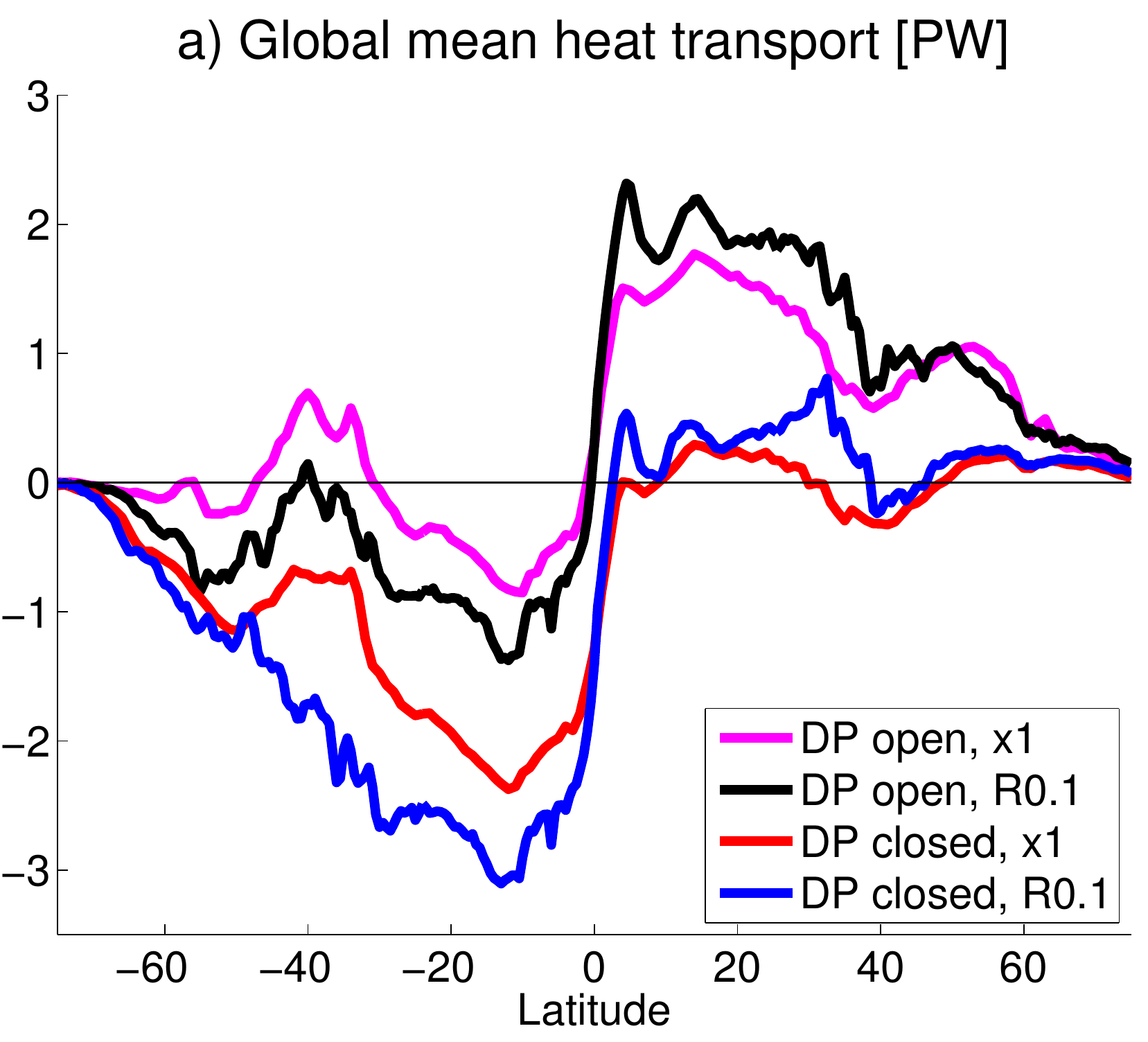}
 \noindent\includegraphics[width=13pc]{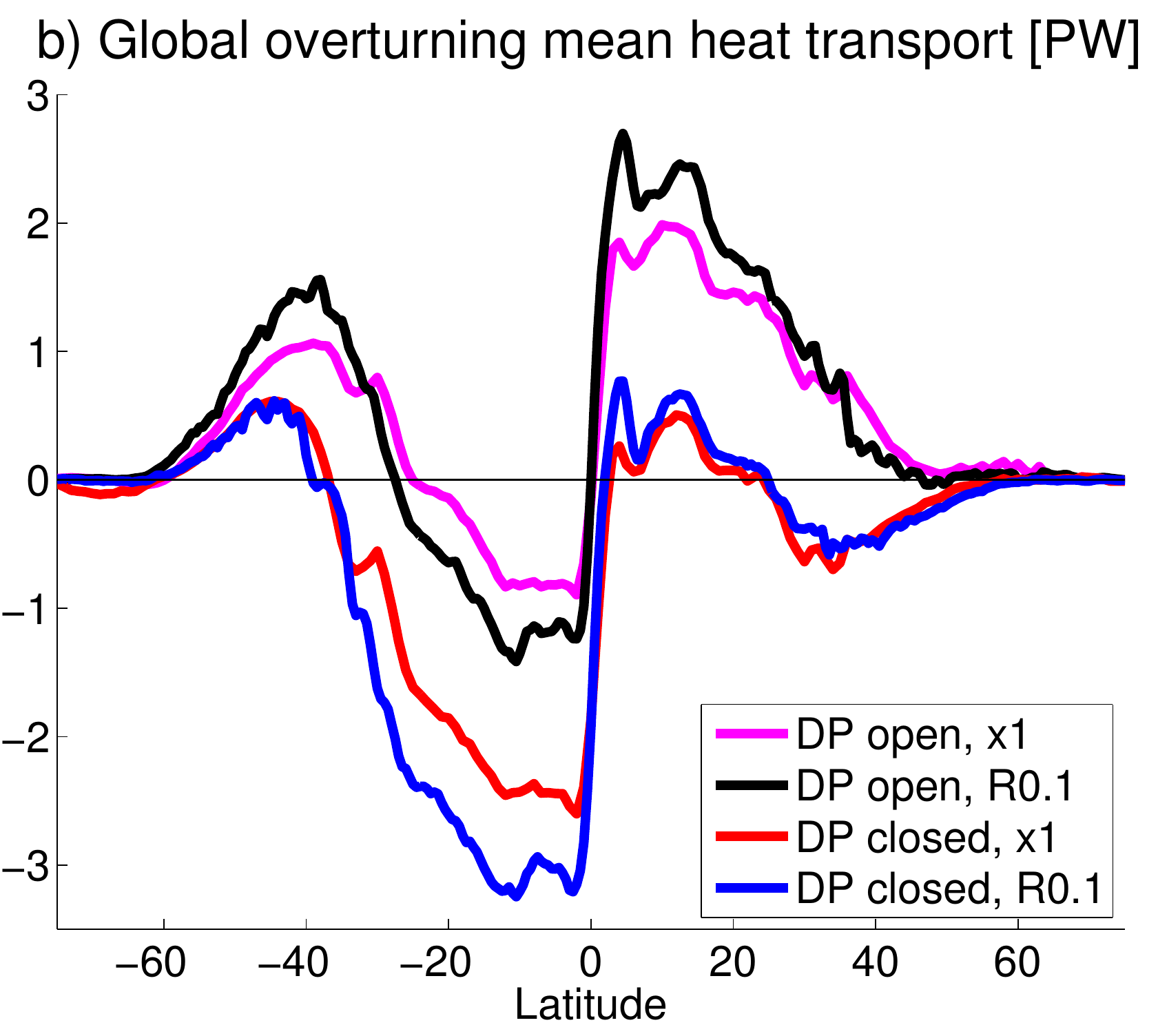}
 \noindent\includegraphics[width=13pc]{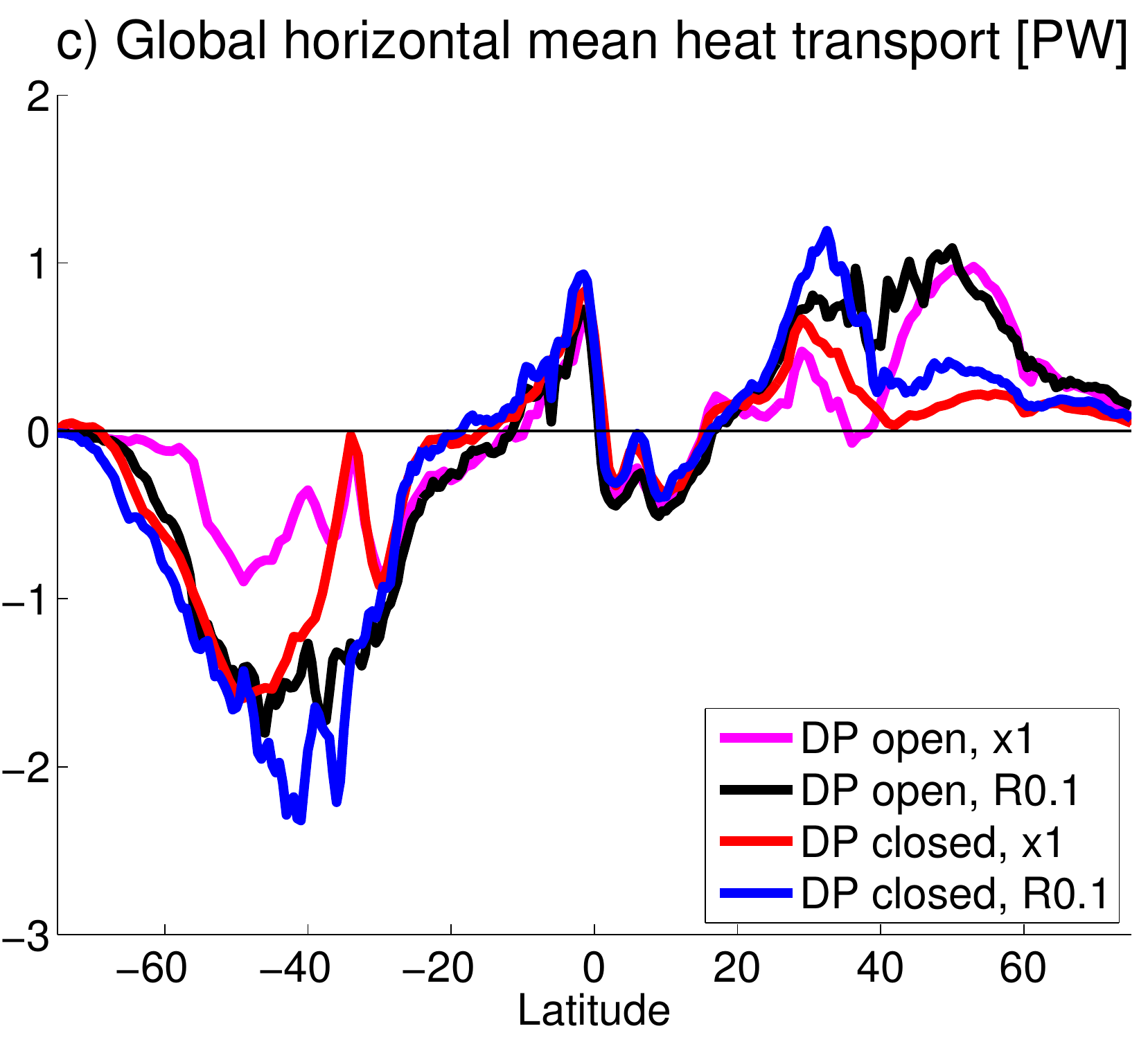}
\caption{The (a) time-mean component of the global advective meridional heat transport [PW]
and its decomposition into components related to (b) meridional overturning circulations
and (c) horizontal circulations are shown for all simulations.
The annual averages are performed for the same years as in Fig. \ref{MHT_SHF1}.
}
\label{MHT_SHF3}
\end{figure}

Figure \ref{MHT_SHF3} shows both $\overline{MHT}_O$ (Fig. \ref{MHT_SHF3}b) and $\overline{MHT}_H$ (Fig. \ref{MHT_SHF3}c) for all simulations.
As before we find that the eddying simulations (blue and black curves) generally show larger magnitudes
than the corresponding non-eddying simulations (red and magenta curves) both in $\overline{MHT}_O$ and $\overline{MHT}_H$.
Moreover, the differences in $MHT$ for both $R0.1$ versus $x1$ as well as $\mathrm{DP^{op}}$ versus $\mathrm{DP^{cl}}$
have a clear latitudinal separation:
Within the tropics differences in $MHT$ are mostly related to meridional overturning circulations (curves of $\overline{MHT}_H$ largely overlap)
whereas in the subpolar regions differences are mostly related to horizontal circulations (curves of $\overline{MHT}_O$ largely overlap or go to zero).

More precisely, similar to \cite{Volkov:10} we find that within the SO for both $\mathrm{DP^{op}}$ simulations (black and magenta curves) the northward $\overline{MHT}_O$ is exceeded by the southward $\overline{MHT}_H$, 
leading to a net southward $MHT$ in the SO\footnote{
\cite{Treguier:07} (see also \cite{Abernathey:14}) compute $\overline{MHT}_O$ and $\overline{MHT}_H$ within the SO across streamlines instead of latitude circles (i.e. for a $\mathrm{DP^{op}}$ simulation).
Their results are similar south of $60^\circ S$ (i.e. dominant $\overline{MHT}_H$ within the region of subpolar gyres)
but indicate that within the ACC belt ($60^\circ S-48^\circ S$) the $\overline{MHT}_O$ (related to Ekman-driven overturning) may be more dominant than $\overline{MHT}_H$.
However, since we compare $\mathrm{DP^{op}}$ simulations with $\mathrm{DP^{cl}}$ simulations (void of circumpolar streamlines) we compute the heat transports only across latitude circles in this study.
}.
For $\mathrm{DP^{cl}}$ (blue and red curves) the $\overline{MHT}_O$ decreases (remaining northward)
whereas the magnitude of $\overline{MHT}_H$ increases leading to the increased southward $MHT$ in the SO.

The previously noted difference in poleward $MHT$ in the SO
between $\mathrm{DP^{op}_{R0.1}}$ and $\mathrm{DP^{op}_{x1}}$ (section \ref{MEMHT0} and Fig. \ref{MHT_SHF1}a) reappears here mainly in $\overline{MHT}_H$
and then nearly vanishes for $\mathrm{DP^{cl}}$.
Consequently, differences between $\mathrm{DP^{op}_{R0.1}}$ and $\mathrm{DP^{op}_{x1}}$ in $MHT$ within the SO
are mostly related to the barotropic circulation, that is,
to different partitionings of the SO flow field into ACC and subpolar gyres (section \ref{barotropic_sec} and Fig. \ref{barotropic}a,c).
In other words, the weaker ACC transport in $\mathrm{DP^{op}_{R0.1}}$ implies
a larger subpolar gyre system in the SO leading to larger poleward $MHT$ compared to $\mathrm{DP^{op}_{x1}}$.
For $\mathrm{DP^{cl}}$ the partitioning of the subpolar flow field is suspended such that the $MHT$s for both configurations $x1$ and $R0.1$ are more similar in the SO.
Ultimately, the larger change in $MHT$ within the $\mathrm{DP_{x1}}$ simulations
induces the larger temperature changes at the Antarctic coast (section \ref{SST_sec} and Fig. \ref{SST}).

Within the tropics $\overline{MHT}_O$ and $\overline{MHT}_H$ also oppose each other.
But here $\overline{MHT}_O$ dominates the $MHT$ and carries the differences between
both $R0.1$ versus $x1$ and $\mathrm{DP^{op}}$ versus $\mathrm{DP^{cl}}$.
We note that in the tropics of the SH the $\overline{MHT}_O$ of $\mathrm{DP^{op}}$ (black and magenta curves)
is southward which indicates that the southward $MHT$ by the shallow wind-driven overturning cells
exceeds the northward $MHT$ by the deep overturning related to NADW formation\footnote{
Note that \cite{Boccaletti:05,Ferrari:11} show that abyssal circulations (related to Antarctic Bottom Water)
do not transport significant heat due to the small temperature differences in the abyss.
More generally, \cite{Ferrari:11} point out that the pathways of mass, represented by the streamfunction,
can be very different from the pathways of heat.}.
Similarly, we find that in the tropics of the NH the $\overline{MHT}_O$ of $\mathrm{DP^{cl}}$ (blue and red curves)
is drastically decreased (due to the shutdown of NADW formation) but still northward.
That is, the northward $MHT$ by the shallow wind-driven overturning cells 
exceeds the southward $MHT$ by the deep diffusively driven overturning shown in Fig. \ref{AMOC}b,d.
In contrast, around $35^\circ N$ the $\overline{MHT}_O$ of $\mathrm{DP^{cl}}$ indeed becomes southward (i.e. is dominated by deep overturning) and the northward subtropical gyre $MHT$ is mostly present in $\overline{MHT}_H$ (i.e. heat transported horizontally).

Finally, in the northern subpolar region $\overline{MHT}_H$ dominates $MHT$ again (i.e. heat is transported horizontally by the subpolar gyre).
However, we note that here the reduction in $\overline{MHT}_H$ due to closing DP is a secondary effect.
Namely the strengthening of $\overline{MHT}_H$ in the SO related to the integration of the ACC into the subpolar gyres
as well as the changes in $\overline{MHT}_O$ related to the shutdown of NADW formation
are direct consequences of closing DP.
In contrast, the northern subpolar gyres are not directly affected by a closed DP
but it is the ceasing heat supply by the deep overturning circulation (seen in $\overline{MHT}_O$)
which subsequently leads to a reduced northward $MHT$ by the North Atlantic subpolar gyre.

\section{Summary and discussion}\label{discussion}
The effects of ocean gateways on the global ocean circulation and climate have been explored
with a variety of general circulation models \citep{Yang:13}.
In particular, SO gateways have been investigated due to their possibly crucial role
in the onset of Antarctic glaciation during the Eocene-Oligcene transition \citep{Kennett:77,Toggweiler:00,Sijp:11}.
However, all of these models employ low model-grid resolutions (larger than $2^\circ$ horizontally) such that the effects of mesoscale 
eddies have to be parameterized. In this study, 
we examined the effects of closing DP within a state-of-the-art high-resolution global ocean model (nominal horizontal resolution of $0.1^\circ$) which allows the explicit representation of energetic mesoscale features 
as well as finer topographic structures. 
For comparison, we also considered the same model experiments at lower non-eddying resolution (nominal $1.0^\circ$). 
The closed DP simulations are configured as sensitivity studies in the sense that
the closure of DP is the only change in the ocean bathymetry which is otherwise identical to the present-day control simulation.

The results are twofold:
On the one hand, the model behavior on the large scale is qualitatively 
similar at both resolutions, but on the other hand, the quantitative details significantly depend on the chosen resolution.
With respect to the qualitative changes induced by closing DP that are similar at both resolutions we find that
the SST significantly increases all around Antarctica with the maximum warming located west of DP.
In terms of flow field the ACC mostly transforms into part of the subpolar gyres
(which hence strengthen and expand)
whereas the subtropical gyre system is largely unaltered.
Moreover, the overturning circulation related to NADW formation shuts down.
A detailed analysis of the MHT reveals that the changes in eddy MHT are relatively small
such that the changes in MHT are largely dominated by the time-mean fields.
Furthermore, it turns out that the warming around Antarctica is mostly determined
by the changes in the horizontal circulation of the SO, namely,
the equatorward expansion of the subpolar gyres which increases the heat transport towards Antarctica.
In contrast, the collapse of the MOC related to NADW formation
dominates the changes in MHT outside of the SO leading to surface cooling in the North Atlantic.

That both configurations ($x1$ and $R0.1$) show qualitatively similar changes in the large-scale ocean circulation
is in accordance with current theories of both the barotropic circulation and the MOC
since closing DP profoundly changes the leading-order zonal-momentum balance in the SO.
More precisely, closing DP inevitably implies
that geostrophy is 
also zonally established 
as leading order balance within the latitudes and depths of DP \citep{Gill:71,Toggweiler:95}.
Hence, the overall pattern of the closed DP large-scale barotropic circulation can be anticipated from linear theory of the large-scale wind-driven gyre circulation \citep{Pedlosky:96,Huber:04,Huber:06}. 
In particular, former ACC streamlines are mostly integrated into the subpolar gyre system
because the line of zero wind-stress curl in the SO is mostly located around $50^\circ S$ (i.e. north of DP).

Moreover, current theories of the overturning circulation divide the MOC into an adiabatic part and a diabatic part \citep{Kuhlbrodt:07,Wolfe:14}, in particular in the SO \citep{Marshall:03,Olbers:05,Ito:08,Marshall:12}.
The pole-to-pole circulation of NADW is considered to be largely adiabatic
such that NADW leaves the diabatic formation region by sliding adiabatically at depth all the way to the SO,
where it is upwelled mechanically along sloping isopycnals by deep-reaching Ekman suction driven by surface westerlies.
The necessary ingredients for the adiabatic pole-to-pole cell are a circumpolar channel subjected to surface westerlies, which allows the wind-driven circulation to penetrate to great depth, and a set of isopycnals outcropping in both the channel and the NH \citep{Wolfe:14}.
Hence, closing DP is one way (next to closing the ``shared isopycnal window") to collapse the adiabatic pole-to-pole cell
as it turns the deep-reaching Ekman suction into shallow gyre Ekman pumping \citep{Toggweiler:95,Toggweiler:98,Toggweiler:00}.
On the other hand, weak mixing can support diabatic overturning cells.
Mixing and wind forcing near the surface can support diabatic surface cells in the tropics and subtropics;
bottom-intensified mixing can support a diffusively driven deep overturning cell associated with bottom water.
These types of diffusive overturning cells are what remains of the AMOC in the $\mathrm{DP_{cl}}$ configuration
at both resolutions.

However, the quantitative details of the ocean circulation (and related heat transports and temperature distributions)
are significantly sensitive to ocean modeling details (in contrast to suggestions by \cite{Huber:04} and \cite{Huber:06}).
Consequently, we find that the warming around Antarctica is substantially larger 
for the non-eddying configuration ($\sim$$5.5^\circ$C) than for the eddying configuration ($\sim$$2.5^\circ$C).
In turns out that this is a consequence of the subpolar mean flow which
partitions differently into gyres and circumpolar current at different resolutions (the DP transport is $159$ ($119$) $Sv$ for $x1$ ($R0.1$)) leading to different heat transports towards Antarctica.
In other words, different representations of mesoscale eddies and topographic details
(inducing different interactions between mean flow, mesoscale eddies and topography)
do not alter the basic principles of the ocean circulation
but can significantly alter quantitative details.

These quantitative details are especially important when it comes to thresholds.
For example, sufficient conditions for the onset of a circumpolar current within the SO are still under debate \citep{Lefebvre:12}.
On the one hand, it is unclear how deep and/or wide an ocean gateway is required for the development
of a significant circumpolar flow.
On the other hand, it is under investigation if a DP latitude band uninterrupted by land represents a
necessary condition for a strong circumpolar current.
Recently, \cite{Munday:15} suggested that an open Tasman Seaway is not a necessary prerequisite
but that only an open DP is necessary for a substantial circumpolar transport.
Support comes from the model simulations of \cite{Sijp:11} which show strong circumpolar flow around Australia
prior to the opening of the Tasman Seaway,
and an additional southern circumpolar route when the Tasman Seaway is unblocked.
This suggests that Antarctic cooling related to the decrease of the subpolar gyres
due to the emergence of circumpolar flow (as seen in this study)
could have been at work twice,
namely, both when DP opened and subsequently when the Tasman Seaway opened.

Closely related to the inception of a circumpolar current
is the partitioning of the resulting subpolar flow field into circumpolar current and subpolar gyre system.
As shown in this study, the stronger the resulting circumpolar flow (i.e. the more former gyre streamlines
become circumpolar) the larger the reduction in southward heat transport.
Hence, this question is closely related to the circumpolar current transport
for which it is known that mesoscale eddies and details of continental and bathymetric geometry are crucial \citep{Rintoul:01,Olbers:04,Kuhlbrodt:12}
but a robust quantitative theory is still lacking\footnote{
For example, \cite{Hogg:14} find that a substantial amount of difference in model sensitivity
may be due to seemingly minor differences in the idealization of bathymetry. 
In their study a key bathymetric parameter is the extent to which the strong eddy field generated in the circumpolar current can interact with the bottom water formation process near the Antarctic shelf.
This result emphasizes the connection between the abyssal water formation and circumpolar transport \citep{Gent:01}
which might be particularily crucial for both
the onset of a circumpolar flow and the partitioning of the subpolar flow field into subpolar gyres and circumpolar current \citep{Nadeau:15}.
} \citep{Nadeau:15}.

Finally, we note that also the MOC is sensitive to ocean modeling details.
This concerns primarily the diffusively driven part of the MOC as it depends on the mixing coefficients
as well as on the numerical diffusion \citep{Kuhlbrodt:07}.
Moreover, as the diffusively driven overturning affects the
ocean density distribution it impacts the shared isopycnal window between the SH and the NH \citep{Wolfe:14}.
Consequently, the diffusively driven overturning may `precondition' the onset of the adiabatic overturning circulation related to NADW formation \citep{Borrelli:14}.

\section{Conclusions and perspectives}\label{conclusions}



\begin{enumerate}
\item From the detailed analysis of meridional heat transports presented in this study,
we conclude that the \emph{ocean gateway mechanism} responsible for changes in sea surface temperature (SST) around Antarctica is based on changes in the horizontal ocean circulation.
It works as follows (compare also the green lines in Fig. \ref{barotropic}a,b):
The onset of a 
circumpolar current within the Southern Ocean (e.g. due to gateway opening)
turns the most northern subpolar gyre streamlines (i.e. the ones carrying most of the heat towards Antarctica)
into circumpolar streamlines (not turning towards Antarctica).
This reduces the subpolar gyre heat transport towards Antarctica and induces cooling near Antarctica.
The opposite happens if a circumpolar current becomes part of the subpolar gyre circulation (e.g. due to gateway closure).

Since the wind-driven gyres are to first order governed by the fundamental Sverdrup balance
and both the basic wind patterns (i.e. high and low pressure systems)
as well as the south-north SST gradient are inevitable
we expect that the ocean gateway mechanism is robust.
That is, we expect it to be qualitatively independent of both 
the ocean modeling details and the coupling to the atmosphere.
What remains is to adequately quantify the changes in SST due to changes in the subpolar ocean circulation,
and to disentangle these from SST changes induced by other mechanisms like $CO_2$ increase or ice growth.

\item The quantification of the ocean gateway mechanism relates to several aspects like
$(i)$ the sufficient conditions for the onset of a circumpolar current within the Southern Ocean,
$(ii)$ the partitioning of the resulting subpolar flow field into circumpolar current and subpolar gyre system,
and $(iii)$ the corresponding changes in meridional heat transports and $SST$.
These aspects are likely highly sensitive to ocean modeling details.

This concerns the dependence on resolution since different representations of both
mesoscale eddies and topographic details can lead to significantly different interactions between mean flow, mesoscale eddies and topography \citep{Hogg:14}.
Furthermore, this concerns the bathymetric details on their own
since paleogeographic reconstructions face substantial uncertainties
due to inaccuracies in proxies or plate tectonic reconstructions \citep{Baatsen:15}.
In particular, the exact latitudes of the Southern Ocean ocean gateways (as well as their relative positioning)
are uncertain which may substantially effect $(i)-(iii)$.
Ideally one would have to access the uncertainty in the details of continental and bathymetric geometry
by considering an ensemble of topographies (reflecting the key uncertainties)
in order to estimate the probability distributions of crucial measures like $(i)-(iii)$.

\item Finally, in order to get a detailed picture of the physical processes actually responsible for the changes
in heat transport and temperature around Antarctica within the different models of paleoclimatic studies,
we suggest more precise attributions of the pathways of heat by decomposing the heat transport into dynamical components,
as performed in this study (as well as e.g. outlined in \cite{Ferrari:11} and applied in \cite{Yang:15}).

Moreover, each tracer (e.g. salt, chemical and organic tracers) follows different pathways in a zonally integrated picture as a result of the different sources and sinks, even though all tracers are transported by the same three
dimensional velocity field \citep{Ferrari:11,Viebahn:12}.
Paleoclimatic studies represent a distinguished context to explore the diversity of tracer pathways.
Hence, we suggest to perform similar decompositions
also for other tracers relevant for Cenozoic climate change
like planktic foraminifera \citep{Sebille:15} or carbon \citep{Zachos:08,Ito:04,Ito:10}.


\end{enumerate}


%
%
%
\section{Additional figures}

%
%
%
%

\begin{acknowledgments}
This work was funded by the Netherlands Organization for Scientific Research (NWO), Earth and Life Sciences,
through project ALW 802.01.024.
The computations were done on the Cartesius at SURFsara in Amsterdam.
The use of the SURFsara computing facilities was sponsored by NWO under the project SH-209-14.
\end{acknowledgments}

%
%
%
%
%
%
%
%
%
{\clearpage}
\bibliographystyle{amseng}
\bibliography{references}
\end{article}
%
%
%
%
%
%
 
\end{document}